\documentclass[12pt,a4paper]{article}
\usepackage[utf8]{inputenc}
\usepackage{lmodern}
\usepackage[T1]{fontenc}
\usepackage[english]{babel}
\usepackage{ifpdf}
\usepackage[usenames,dvipsnames]{xcolor}
\usepackage{graphicx}
\usepackage[shortlabels]{enumitem}
\usepackage[numbers,sort]{natbib}
\usepackage[font=footnotesize,width=\textwidth]{caption}
\usepackage{sectsty}
\usepackage{tcolorbox}
\usepackage{hyperref}
\usepackage{uri}
\usepackage[top=1.8cm, bottom=2.8cm]{geometry}
\usepackage[affil-it]{authblk}
\usepackage[parfill]{parskip}
\usepackage[setbuildercolon]{matmatix}

\ifpdf
  \hypersetup{
    pdftitle={Mathematically tractable models of random phylogenetic networks: an overview of some recent developements}
  }
\fi

\hypersetup{colorlinks, allcolors = {MidnightBlue}}

\allsectionsfont{\boldmath}

\newenvironment{mathlist}
{\begin{enumerate}[label={\upshape(\roman*)}, align=left, widest=iii, leftmargin=*]}
{\end{enumerate}\ignorespacesafterend}

\makeatletter
\renewcommand*{\defas}{%
\mathrel{\rlap{\raisebox{0.3ex}{$\m@th\cdot$}}\raisebox{-0.3ex}{$\m@th\cdot$}}=}
\newcommand*{\asdef}{%
\mathrel{{=\llap{\raisebox{0.3ex}{$\m@th\cdot$}}\llap{\raisebox{-0.3ex}{$\m@th\cdot$}}}}}
\makeatother

\makeatletter
\newtheorem*{rep@theorem}{\rep@title}
\newcommand{\newreptheorem}[2]{%
\newenvironment{rep#1}[1]{%
 \def\rep@title{#2 \ref*{##1}}%
 \begin{rep@theorem}}%
 {\end{rep@theorem}}}
\makeatother

\newreptheorem{theorem}{Theorem}

\newreptheorem{proposition}{Proposition}

\newreptheorem{lemma}{Lemma}

\theoremstyle{definition}

\newtcolorbox[auto counter]{mybox}[2][]{%
    floatplacement=htbp,
    float,
    sharp corners=all,
    boxrule=1pt,
    titlerule=-1pt,
    left=1ex,
    right=1ex,
    toptitle=1ex,
    colbacktitle=black!5!white,
    coltitle=black,
    fonttitle=\footnotesize\bfseries,
    fontupper=\footnotesize,
    title=Box~\thetcbcounter: #2,
    #1}

\newcommand{\SuppMat}{\hyperref[secSuppMat]{Supplementary Material}}

\title{\vspace{-1.3cm}Mathematically tractable models of\linebreak random phylogenetic networks: %
an overview of some recent developments}

\begin{document}

\author{François Bienvenu}

\renewcommand\Affilfont{\itshape\small\renewcommand{\baselinestretch}{1}}

\affil{Université de Franche-Comté, CNRS, LmB, F-25000 Besançon, France.\vspace{1ex}}

\maketitle

\begin{abstract}
Models of random phylogenetic networks have been used since the
inception of the field, but the introduction and rigorous study of
mathematically tractable models is a much more recent topic that has gained
momentum in the last 5~years.
This manuscript discusses some recent developments
in the field through a selection of examples.
The emphasis is on the techniques rather than on the results themselves,
and on probabilistic tools rather than on combinatorial ones.
\end{abstract}

\setcounter{tocdepth}{2}
\tableofcontents

\pagebreak

\section{Introduction} \label{secIntro}

Over the last two decades, phylogenetic networks have become one of the most
active areas of research in mathematical phylogenetics.
Much of the work so far has consisted in identifying relevant concepts
and formally defining classes of networks; then in understanding the
relationships between these classes and the mathematical properties of each
class (with a special focus on algorithmic properties and enumeration).
For accounts of the development and current state of the field, see e.g.\ the
classic textbooks \cite{huson2010phylogenetic}
and~\cite{gusfield2014recombinatorics}; the book
chapters~\cite[Chapter~10]{steel2016phylogeny} and
\cite[Chapter~12]{zhang2019clusters}; and the recent review~\cite{kong2022classes}.

Recently -- starting for the most part in the early 2020's -- a new
line of research has started to emerge: the introduction and mathematical study
of models of random phylogenetic networks.
Of course, random phylogenetic networks are not, in themselves, a new topic.
In fact, the recognition of their importance and the formulation of
models for use in concrete applications began in the early stages of the
development of phylogenetic networks, and predate the formal introduction of many of
today's prominent classes of networks: see for instance
\cite{nakhleh2002towards,morin2006netgen}.  What is new, however, is the
introduction of models of random phylogenetic networks that are designed with
the specific goal of being studied mathematically -- what we call here
\emph{tractable models}.

Besides their intrinsic mathematical interest, there are good reasons
to study tractable models of random phylogenetic networks: first,
having models with few parameters and whose behavior we understand thoroughly 
is useful -- sometimes crucial -- in concrete applications.
A prime example of this is the development of Bayesian approaches to
reconstruct phylogenetic networks from data, such as~\cite{zhang2018bayesian},
because one needs a prior for the network.
Maximum likelihood estimation is not exempt from the need for tractable models,
because in situations where the data co-evolves with the network,
a good mathematical understanding may be needed to compute the
likelihood of a network.
Other examples of concrete applications include the development of rigorous
statistical tests and the validation of reconstruction methods
(for the latter, mathematical tractability is not vital,
provided the model can be simulated efficiently; but a thorough understanding
of the model can be helpful for this, see Section~\ref{secEnumAndSampling}
below).

Second, a more conceptual -- and more disputed -- interest of
tractable models is that they are valuable for the comprehension of
natural phenomena by the human mind.
Indeed, if we could define and simulate efficiently a model that perfectly
describes, say, the evolution of species, we could use it to fully characterize the
effect of each parameter, and then describe those effects in human terms.
That would make us very knowledgeable about the evolution of species; but that
would not necessarily mean that we \emph{understand} it.
By simplifying the model down to a level where its behavior can be explained by
arguments whose every step can be rephrased in logical terms and checked to be
correct, we access a different, arguably deeper level of understanding.
The inevitable price to pay for this additional understanding is
that effects and parameters have to be discarded, which
makes the model a less accurate description of reality.
But this is not necessarily a problem, since from that perspective the true
goal of tractable models is not to accurately describe natural phenomena: it is
to shed a specific light on them that contributes to the multiple facets of our
understanding.

In practice, the concept of tractability is relative: whether a model is
considered tractable depends on its intended use.
Because networks are complex objects and have a high information content, they often
need to be studied through the lens of summary statistics that
capture specific aspects of their structure and geometry.
These could be simple quantities, such as pattern counts
(number of reticulations, number of cherries); or more elaborate ones
that aim at quantifying ``intuitive'' notions,  such as 
phylogenetic balance (detailed in Section~\ref{secExampleStats}).
However, instead of directly focusing on specific statistics, it is sometimes
more efficient to study objects that encode all or part of
the network, and from which multiple statistics can be recovered -- similar
to contour processes for trees (see e.g.~\cite{lambert2017probabilistic}).
Related to this are two notions of \emph{limits of networks} that can be used to
characterize the geometry of large networks on different scales.
Although these notions have become part of the standard probability toolbox,
so far their use has been limited in the field of phylogenetic networks.

In the rest of this document, we focus exclusively on networks that aim
to give an explicit description of the evolutionary history of a set of
taxa -- so-called \emph{explicit networks} \cite{huson2010phylogenetic,
gusfield2014recombinatorics,steel2016phylogeny,kong2022classes}.
Because we are interested in the mathematical study of models,
several interesting studies on random phylogenetic networks are not discussed,
such as the recent simulation studies \cite{janssen2021comparing} and
\cite{justison2023exploring}.
Importantly, the goal of this article is not to give an exhaustive review of
existing models and results. Instead, it is to present a selection of
techniques that have recently proved useful to study tractable models.
The focus is on probabilistic tools -- however this should not
create the false impression that these are the main tools that have been
used successfully to study random phylogenetic networks:
it is precisely because most of the existing work is based on combinatorial
approaches that this article choses to focus on techniques that appear to be
lesser-known in the field.
In order to make the manuscript accessible, the level of mathematical formalism
is intentionally kept to a minimum, and only the main ideas are sketched.
However, references to in-depth presentations are given for each topic.

The rest of the manuscript is organized as follows: Section~\ref{secModels}
introduces two simple models of random phylogenetic networks that are then used
as examples throughout the rest of the text: birth-hybridization processes, and
uniform galled trees.
After recalling the definition of some summary statistics,
Section~\ref{secSummaryStats} gives a brief overview of how the method of
moments has been used to study their asymptotic distribution in large random
phylogenetic networks; it then presents a less computation-intensive
alternative: the Stein--Chen method.
Next, in Section~\ref{secBlowups} we introduce an important class of models, of
which uniform galled-trees are a specific instance and for which many powerful
branching process techniques can be used: blowups of conditioned Galton--Watson
trees.
Finally, Section~\ref{secLimits} gives a short introduction to the notions of
local limit and Gromov--Hausdorff limit, with examples showing how these can be
used to answer concrete questions concerning blowups of conditioned
Galton--Watson trees.
Some concluding comments and perspectives are given in Section~\ref{secConclusion}.

\section{Examples of tractable models} \label{secModels}

The goal of this section is to introduce some vocabulary and some models that
will then be used throughout the rest of the text.
We only present two models: the birth-hybridization process, and uniform galled
trees (with a brief discussion of the more general uniform level-$k$ networks).
The reason for choosing to focus on these specific models is that they are very
natural and among the simplest models of random phylogenetic networks that
retain some mathematical interest; and that as of today they are arguably the
models whose behavior we understand best.

Note however that these are not the only tractable models of random
phylogenetic networks that have been discussed in the literature:
Section~\ref{secMoments} contains several other examples of models
that have been studied using combinatorial methods.
In addition to this, some process-based models that have been studied
through simulations should be somewhat tractable.
These include some of the tree-based models used in
\cite{janssen2021comparing}, and possibly the model generating level-$k$ LGT
networks introduced in~\cite{pons2019generation}, at least for $k = 1$.
Finally, because in the model underlying~\cite{justison2023exploring} the
hybridization rate becomes exactly equal to~0 once lineages have diverged past
a certain threshold, it should be possible to apply some of the
branching-process techniques presented in Sections~\ref{secBlowups}
and~\ref{secLimits} to study that model.

\subsection{Vocabulary} \label{secVocab}

Recall that we restrict ourselves to explicit networks -- that is, to networks
that give a direct description of the evolutionary history of a set of taxa.
In that context, a phylogenetic network is a rooted connected DAG (directed
acyclic graph). By \emph{rooted}, we mean that there exists a unique vertex
with indegree~0, called the \emph{root}; and by \emph{connected}, we mean
that for every vertex $v$ there is a directed path going from the root to~$v$.
In anticipation of Section~\ref{secLocalLim}, where infinite phylogenetic
networks will play an important role, we allow
networks with a countably infinite number of vertices.
However, we require that these networks be \emph{locally finite} -- meaning
that every vertex has a finite degree.

Note that most formal definitions of phylogenetic networks exclude multi-edges
(i.e.\ having more than one edge between two vertices) and vertices with
indegree and outdegree both equal to~1 -- see
e.g.~\cite{steel2016phylogeny,zhang2019clusters}.
While this is very natural from a biological point of view (and important
in the definition of combinatorial classes), there is no real need
to exclude multi-edges and vertices with in- and outdegree~1 for our purposes,
and allowing them can sometimes be useful.

Let us briefly recall some basic terminology:
if there is an edge from $u$ to $v$, then we say that $u$ is a
\emph{parent} of $v$ and that $v$ is a \emph{child} of $u$. Similarly, if there
is a directed path from $u$ to $v$ then we say that $u$ is an \emph{ancestor}
of $v$ and that $v$ is a \emph{descendant} of $u$.
Vertices with at most one parent are called \emph{tree vertices}, and vertices
with more than one parent are called \emph{reticulations}. Correspondingly,
the edge $u \to v$ is called a \emph{tree edge} if $v$ is a tree vertex and
a \emph{reticulation edge} if $v$ is a reticulation.
Finally, vertices with no children are called \emph{leaves}, and
non-leaf vertices are called \emph{internal vertices}.

\break

\begin{figure}[h]
  \centering
  \captionsetup{width=.95\linewidth}
  \includegraphics[width=0.8\linewidth]{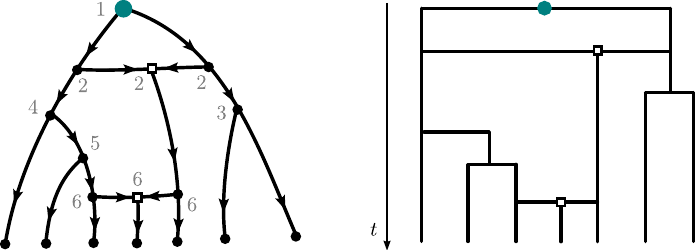}
  \caption{Left, an example of a temporal binary phylogenetic network, along
  with a temporal labeling of its internal vertices.  The root is the green
  dot, and the white squares indicate reticulations. Note that this representation
  of the network is redundant: for instance, the direction of the edges can be
  deduced from the position of the root and of the reticulations. Thus,
  in this document we will frequently use simplified representations, such as
  the one shown on the right. Right, a time-embedding
  of the network on the left that is compatible with its temporal labeling,
  with time on the vertical axis.}
  \label{figExampleNetwork}
\end{figure}

From a biological point of view, it is unlikely that a single speciation event
could result in the creation of more than two species, or that more than
two lineages could combine to form a new one. As a result, phylogenetic
networks that rule out these events are of particular interest.
A phylogenetic network is said to be \emph{binary} if its root has outdegree~2
and if every other internal vertex has either:
\begin{itemize}
  \item indegree~1 and outdegree~2 (tree vertices);
  \item indegree~2 and outdegree~1 (reticulations).
\end{itemize}
Likewise, explicit phylogenetic networks are supposed to be the direct products
of time-embedded evolutionary processes.
This imposes constraints on their structure that persist even after the
time-embedding is lost.
These structural constraints can be blurred by subsampling, to the point of
vanishing altogether; but if a network contains all ancestors of the taxa
corresponding to its leaves and if the evolutionary events represented by
the various elements of the networks can be treated as instantaneous, then the
constraints take a simple form.
This motivates the following definition, due to~\cite{baroni2006hybrids}:
a phylogenetic network is said to be \emph{temporal} or \emph{time-consistent}
if there exists a real-valued function $t$ on its vertex set such that, for
every edge $u\to v$:
\begin{mathlist}
\item  $t(u) < t (v)$ if $u\to v$ is a tree edge;
\item  $t(u) = t (v)$ if $u\to v$ is a reticulation edge.
\end{mathlist}
In phylogenetics, such a function $t$ is usually called
\emph{temporal labeling} \cite{kong2022classes, steel2016phylogeny,
zhang2019clusters} or a \emph{time-stamp function} \cite{huson2010phylogenetic}.
Here, in keeping with the random tree community, we will refer to
it as a \emph{time-embedding} and we will use the term
\emph{time-embedded network} to refer to a temporal phylogenetic network
together with its temporal labeling.

As a final point regarding vocabulary, note that in the rest of this document
we will frequently use the terms \emph{combinatorial models} 
to refer to models that correspond, by definition, to the uniform distribution
on some well-specified combinatorial class of networks; and
\emph{process-based models} for models that result from time-embedded processes.
However, this distinction is not always relevant, because some models
fall in both categories -- similar to how the Yule tree shape can be obtained
both from the Yule process and by sampling uniformly at random from the set of
ranked binary trees with a fixed number of labeled leaves. In fact, being able
to switch from one view point to the other is often what makes a model
tractable; we will see several examples of this.

\enlargethispage{3ex}
\vspace{-0.2cm}

\subsection{Birth-hybridization processes and RTCNs} \label{secRTCNs}

\begin{figure}[h]
  \centering
  \captionsetup{width=0.9\linewidth}
  \includegraphics[width=0.8\linewidth]{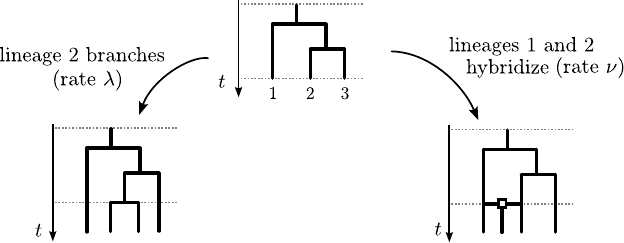}
  \caption{Illustration of the two types of events (branching, at rate
  $\lambda$ per lineage; and hybridization, at rate $\nu$ per pair of lineages)
  governing the dynamics of the birth-hybridization model.}
  \label{figExampleBirthHybridization}
\end{figure}

One of the simplest models of phylogenetic network generated
by a time-embedded process is the following: starting from a single
lineage at time~$0$, let
\begin{itemize}
  \item each lineage branch at constant rate $\lambda$ --
    when a lineage branches, it splits into two new lineages
    (see Figure~\ref{figExampleBirthHybridization});
  \item each pair of lineages hybridize at constant rate $\nu$ --
    when two lineages hybridize, they both survive and a new lineage
    is created with both of them as parents 
    (see Figure~\ref{figExampleBirthHybridization}).
\end{itemize}

Thus, the model is a continuous-time Markov chain where, conditional on there
being $k$ lineages, the time to the next event is an exponential variable
with parameter $k\lambda + \smash{\scriptstyle\tbinom{k}{2}} \nu$. That event
is then chosen uniformly at random among the $k$ possible branching events and
$\smash{\tbinom{k}{2}}$ possible hybridizations.

This model was first described in \cite{zhang2018bayesian}, where it served
as a prior for the Bayesian inference of species networks from multilocus
sequence data.
It was also used in the simulation study \cite{janssen2021comparing}, where it
is referred to as the ``ZODS model''.
The case $\nu = 2 \lambda$ is of particular interest from
a mathematical point of view, because in that case the model produces
phylogenetic networks that are uniformly distributed on the set of
leaf-labeled ranked tree-child networks.
Ranked tree-child networks (RTCNs) are a combinatorial class of
phylogenetic networks recently introduced in \cite{bienvenu2022combinatorial}.
They can be viewed as an extension of ranked binary trees that allows for
hybridizations, and can be succinctly described as ``tree-child networks
embedded in discrete time''.
More specifically, RTCNs are temporal tree-child networks
endowed with an additional structure called the \emph{ranking}.
As for trees, the ranking is a strict ordering of internal vertices that is
compatible with the genealogical order.
However, whereas for trees it takes the form of a strict total order on the
set of all internal vertices (see
e.g.~\cite[Section~2.3]{semple2003phylogenetics}), in the case of tree-child
networks, only some internal vertices are ordered: the tree vertices with no
reticulation children (which correspond to branching events); and the
reticulations (which correspond to hybridizations).

\begin{mybox}[label=boxLinearHybridization]{the birth-hybridization process with linear hybridization rate}
To address the fact that the standard birth-hybridization model produces
networks with too many reticulations, it is possible to change the
hybridization rate and make it an arbitrary function of the number of
lineages.
In particular, making the per-lineage hybridization rate constant
is a natural choice, as this could reflect situations where the number of
lineages with which each lineage can hybridize is bounded (and thus
can be assumed to be, on average, constant in time).
This limit on the number of potential partners for hybridization could be
interpreted as resulting from geographical constraints or from
steric hindrance in some abstract space where the lineages are located
(such as a phenotypic space).\smallskip

By \emph{birth-hybridization model with linear hybridization rate}, we refer
to the birth-hybridization model where each lineage branches at constant
rate~$\lambda$ and -- when there are at least two lineages --
choses a lineage with which to hybridize uniformly at random at rate~$\eta$.
Thus, starting from two lineages,
the total number of lineages
is described by a Yule process with parameter $\lambda + \eta$, and
the network can be sampled by generating a Yule tree and then adding
horizontal edges -- independently and with probability
$\theta \defas \eta / (\lambda + \eta)$ for each tree vertex (except the
root). \smallskip

Although the structure of the networks produced by this model differs greatly
from that of uniform leaf-labeled RTCNs and of networks produced by the original
birth-hybridization model of~\cite{zhang2018bayesian}, many of the arguments
used in~\cite{bienvenu2022combinatorial} to study RTCNs
still apply.
This is because (1)~the profile of the network -- i.e.\ the
sequence $(q_i)_{i\geq 1}$ such that $q_i = 1$ if the $i$-th event in time is
a branching and 0 otherwise -- consists of independent Bernoulli variables with
parameter~$1 - \theta$, except for $q_1 = 1$ (thus, the number of
reticulations in a network with $n$ leaves is a binomial
variable with parameters $n-2$ and $\theta$); and (2)
stopped upon reaching $n$ vertices and conditioned to have profile
$\mathbf{q} = (q_1, \ldots, q_{n-1})$, the network is distributed
as a uniform leaf-labeled RTCN with profile~$\mathbf{q}$.\smallskip

From this observation, measures of the typical distance between the root
and the leaves, can be obtained: let $\gamma_n^{_\downarrow}$ be the random
path obtained by starting from the root and following the edges of the network
until we reach a leaf, respecting the orientation of the edges and choosing
each outgoing uniformly at random and independently of everything else when
we reach a tree vertex. 
Let $\mathrm{length}(\gamma_n^{_\downarrow})$ denote
the number of tree edges of $\gamma_n^{_\downarrow}$,
with the idea -- as in the definition of temporal networks --  that
reticulation edges correspond to instantaneous hybridization events.
Then, following the proof of \cite[Theorem~5.2]{bienvenu2022combinatorial} --
see the \SuppMat{} -- we get
\begin{equation} \label{eqLengthGammaDown}
  \mathrm{length}(\gamma_n^{_\downarrow}) \;\overset{d}{=}\;
  1 \;+\; \sum_{i = 2}^{n-1} X_i \,, 
\end{equation}
where the random variables $X_2, \ldots, X_{n-1}$ are independent with
$X_i \sim \mathrm{Bernoulli}((1+\theta) / i)$.
Thus, $\ell^{_\downarrow}_n \defas
\Expec{\mathrm{length}(\gamma_n^{_\downarrow}}) = (1 + \theta) H_{n-1} - \theta$,
where $H_n$ denotes the $n$-th harmonic number; and using, e.g., the
Stein--Chen method (see Section~\ref{secSteinChen}) we get
\begin{equation}
  d_{\mathrm{TV}}\big(\mathrm{length}(\gamma_n^{_\downarrow}), \,
  \mathrm{Poisson}(\ell_n^{_\downarrow})\big) \;\tendsto{n\to\infty}\;  0\,, 
\end{equation}
where $d_{\mathrm{TV}}$ denotes the total variation distance. In particular,
this implies the following central limit theorem:
\begin{equation}
  \frac{\mathrm{length}(\gamma_n^{_\downarrow}) - (1 + \theta)
  \log n}{\sqrt{(1 + \theta) \log n}}
  \;\tendsto[d]{n\to \infty} \; \mathcal{N}(0, 1) \,, 
\end{equation}
where $\mathcal{N}(0, 1)$ denotes the standard normal distribution.\smallskip

A corresponding result can be obtained for the random path $\gamma_n^{_\uparrow}$
obtained by choosing a leaf uniformly at random and then following the edges in
opposite direction until we reach the root, choosing each incoming edge
uniformly at random and independently of everything else when we reach a
reticulation. This time, we get
$\ell^{_\uparrow}_n = (2 + \theta) (H_n - \tfrac{3}{2}) + 1$;
see the \SuppMat.
\end{mybox}

The dual nature of uniform leaf-labeled RTCNs as combinatorial and process-based
models makes them of specific interest from a mathematical point of view.
In addition to the birth-hybridization process with $\lambda = 2\nu$, there are
several ways to sample them, such as using a coalescent process
\cite{bienvenu2022combinatorial} and from other uniformly distributed
combinatorial objects -- including trees, permutations and even
river-crossings~\cite{caraceni2022bijections}.
Besides simple quantities such as the number of reticulations or the number
of cherries (that is, tree vertices whose children are both leaves), 
tractable statistics include measures of the typical distance between the root
and the leaves, as quantified by the lengths of various random paths
\cite{bienvenu2022combinatorial}; and the number
of occurrences of various fringe patterns, including complex
ones~\cite{fuchs2024limit}.
With the exception of the sampling using bijections
with other combinatorial classes, most of these results can
be generalized to the birth-hybridization model for arbitrary values of
$\lambda > 0$ and $\nu > 0$; the formulas become more complex, but the
arguments are the same.

Despite its natural formulation, and for all its mathematical interest,
it is important to note that from a biological point of view
the birth-hybridization model has two major flaws:
\begin{itemize}
  \item the fact that the overall hybridization rate increases quadratically,
    which entails that there are too many reticulations
    and that the number of lineages explodes (i.e.\ becomes infinite in
    finite time);
  \item the fact that all pairs of lineages hybridize at the same rate: instead,
    more closely related lineages should hybridize at a higher rate.
\end{itemize}
Of course, these two points are closely linked, since decreasing the
hybridization rate of phylogenetically distant lineages would decrease the
overall hybridization rate.
However -- and even though this is not entirely satisfactory from
a modelling point of view -- it is possible to tackle the first point
separately, by making the overall hybridization rate an arbitrary function
of the number of lineages; see Box~\ref{boxLinearHybridization}.
We come back to the challenges posed by the second point in
Section~\ref{secConclusion}.

\subsection{Galled trees and level-$k$ networks} \label{secGalledTrees}

Galled trees are one of the oldest classes of phylogenetic networks
to have been formally described \cite{gusfield2004fine, gusfield2004optimal}.
They play an important role in the field of phylogenetic networks,
particularly in algorithmics; see \cite[Chapter~8]{gusfield2014recombinatorics}.
Before defining them, we need to introduce some vocabulary: a
\emph{reticulation cycle} is a pair of edge-disjoint directed paths
that start in a common vertex~$u$, called the root of the cycle,
and end in a common vertex~$v$, called its reticulation. 
In a phylogenetic network, each reticulation belongs to at least one
reticulation cycle.

A (binary) \emph{galled tree} is a binary phylogenetic network without
multi-edges such that every reticulation belongs to exactly one
reticulation cycle and the reticulation cycles are disjoint.
In the rest of this document, we focus exclusively on binary
galled trees and refer to them simply as ``galled trees''.

\begin{figure}[h]
  \centering
  \captionsetup{width=.95\linewidth}
  \includegraphics[width=0.8\linewidth]{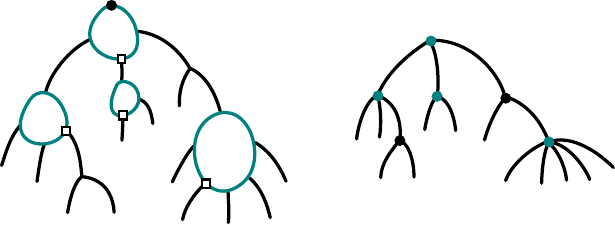}
  \caption{Left, a galled tree. The black dot on top is the root and
  the white squares indicate reticulations. The edges are directed, but
  as previously their orientation is not shown as it is imposed by the
  position of the reticulations. This galled tree can be obtained from
  the tree on the right by replacing the internal vertices highlighted in
  green by reticulation cycles. Galled trees are exactly the 
  phylogenetic networks that can be built in such a way.}
  \label{figGalledTree}
\end{figure}

Another way to think about galled trees -- which is just a trivial
reformulation of their definition -- is that they are exactly the binary
phylogenetic networks that can be obtained by starting from a (non-binary) tree
and then replacing some of its internal vertices by reticulation cycles,
as illustrated in Figure~\ref{figGalledTree}.
The change of viewpoint behind this simple rewording is the key for using the
tools presented in Sections~\ref{secBlowups} and~\ref{secLimits}.

Galled trees have the following natural generalization: recall that a
\emph{biconnected component} of a phylogenetic network -- also known as a
\emph{block} -- is a maximal biconnected subgraph (meaning that
it cannot be disconnected by removing a vertex, and that it is not contained
in a larger subgraph having this property).
The \emph{block-cut tree} of a phylogenetic network is the tree whose vertices
are the blocks of the networks and where two blocks are connected if and only
if they share a vertex (such vertices are known are \emph{cut-vertices}).
A \emph{level-$k$} network is a (binary) phylogenetic network whose blocks have
at most $k$ reticulations. Thus, trees correspond to level-$0$ networks, and
galled trees to level-$1$ networks.

In the context of explicit networks, one of the most natural and best studied
combinatorial class of level-$k$ networks are leaf-labeled level-$k$
networks. For $k =1$ and $k = 2$, they have been enumerated explicitly in
\cite{bouvel2020counting}, using analytic combinatorics; as we will see
below, their asymptotic enumeration for arbitrary values of $k$ was recently
solved in \cite{stufler2022branching}, using branching process techniques.

\enlargethispage{2ex}

Given the importance of leaf-labeled level-$k$ networks in the phylogenetic
networks literature, particularly in the case $k =1$ of galled trees, a natural
model of random phylogenetic network consists in considering the uniform
distribution on the set of level-$k$ networks with $n$ labeled leaves.
This model has been studied extensively by Stufler
in~\cite{stufler2022branching}. This paper  introduced many of the branching
process techniques discussed in this review (couplings with conditioned
Galton--Watson trees, local limits and Gromov--Hausdorff limits) to the field
of phylogenetics networks. In it, Stufler realized that the
structure of a level-$k$ network is in great part determined by the structure of
its block-cut tree, and that this has concrete implications for the
enumeration and the large-scale geometry of uniform level-$k$ network.
For instance, this made it possible to show that the number $a^{(k)}_n$ of
level-$k$ networks with $n$ labeled leaves satisfies
\begin{equation}
  a^{(k)}_n \;\sim\; c_k\, n^{-\frac{3}{2}} \rho_k^{-n}\, n! 
  \qquad \text{as } n\to\infty\,, 
\end{equation}
where the constants $c_k$ and $\rho_k$ have a clear probabilistic
interpretation (and are known explicitly for $k \leq 2$ thanks to the
work of \cite{bouvel2020counting}).

\section{Studying summary statistics} \label{secSummaryStats}

In this section, we present general methods that can be used to study summary
statistics of random phylogenetic networks. Of course, it would not be possible
to give an overview of the whole range of tools that can be used for this,
as this would mean writing a probability textbook.
Therefore,
we only focus on the two methods that have found the most applications so far:
the method of moments and the Stein--Chen method. But first, let us say a
word about summary statistics of phylogenetic networks.

\subsection{Summary statistics of phylogenetic networks} \label{secExampleStats}

In full generality, a summary statistics can refer to any function that assigns
a real number to a phylogeny. In practice, summary statistics
are typically label-invariant and aim at quantifying some specific aspect of
the structure of the phylogeny; hence the fact that they are also referred
to as \emph{shape statistics}.

As mentioned in the introduction, summary statistics play an important role in
phylogenetics, be it for the development of quantitative methods or simply to
inform our mental representation of phylogenies. Yet, in sharp contrast to
phylogenetic trees, for which multitude of summary statistics have been defined
and are routinely used, comparatively few summary statistics are available
for phylogenetic networks.

Many of the statistics used in practice are based on \emph{pattern counts},
i.e.\ on the number of occurrences of specific subgraphs. These patterns can
correspond to a fixed subgraph (e.g, leaves, reticulations, cherries,
reticulated cherries); or to classes of subgraphs (e.g, reticulation cycles,
blocks, shortcuts). Other commonly used statistics
include measures of the \emph{depth} or \emph{height} of a network
(e.g, by the maximum over the leaves of the length of a shortest path from the
root to that leaf); and the \emph{level} of a network, defined as the
maximum number of reticulations in a block.
Statistics such as the number of ways to label or to rank a network have been
studied mathematically, but are rarely used in concrete applications since
they are not always biologically relevant and can be challenging to compute.

In addition to this, a special class of summary statistics deserves its own
treatment: indices of phylogenetic balance, or \emph{balance indices} for 
short.
For phylogenetic trees, balance indices are arguably the most important and
widely used class of summary statistics.
Simply put, they attempt to quantify the idea that some trees
``look more balanced'' than others when we draw them; and even though
there have been attempts at formalizing the notion
\cite{fischer2023tree, lemant2022robust}, in practice a balance index can be
any useful statistic that conforms to our intuitive idea of what it
means for a tree to be ``balanced''.
See \cite[Chapter~33]{felsenstein2004inferring} for a short introduction to
tree balance indices, and
\cite{fischer2023tree} for a comprehensive survey that includes 19 indices.

Despite the profusion of balance indices for trees, as of today there are only
two balance indices (or rather classes of balance indices)
for phylogenetic networks: extensions of the Sackin index; and the $B_2$
index.

\subsubsection{The Sackin index and its extensions to networks} \label{secSackin}

The Sackin index of a tree is a classic balance index that
was introduced by Shao and Sokal in~\cite{shao1990tree}, based
on an idea suggested by Sackin in~\cite{sackin1972good}.
It is defined, for a tree~$T$, as
\begin{equation} \label{eqDefinitionSackin}
  \mathrm{Sackin}(T) \;=\;
  \sum_{\ell \in L} \delta_\ell \,, 
\end{equation}
where the sum runs over the leaves of the tree and $\delta_\ell$ denotes
the \emph{depth} of leaf~$\ell$, i.e.\ its distance to the root.
In computer science, the Sackin index is known as the
\emph{external path length}, see e.g.~\cite[Section~2.3.4.5]{knuth1997art},
and is used as a measure of efficiency for certain
algorithms. Although it is not immediately clear from
its definition that the Sackin index should be related to our intuitive
idea of what it means for a tree to be balanced,
this turns out to be case in practice: the larger the Sackin index, the less
balanced the tree. 
Today, the Sackin index has become the prototypical example of a balance
index.

The definition of the Sackin index given in Equation~\eqref{eqDefinitionSackin}
can be extended to phylogenetic networks, provided we specify what is meant
by the depth $\delta_\ell$ of a leaf: indeed, because there can be more
than one path going from the root to a leaf, the notion of depth is ambiguous
in networks.
Two natural definitions consist in considering (1)~the length of shortest paths
and (2)~the length of longest paths (other possibilities include using the
expected length of random paths, as in Box~\ref{boxLinearHybridization}).
Note however that although these various definitions provide mathematically
natural extensions of the Sackin index of a tree, it is not clear why the
resulting summary statistics should be considered balance indices.

The first study of the Sackin index of a random phylogenetic network
is~\cite{zhang2022sackin}, where the length of longest paths from the root is used
for the depth. The main result of this paper is that, as $n$ goes to infinity,
the expected value of the Sackin index of a phylogenetic network sampled
uniformly at random on the set of simplex networks with $n$ labeled leaves grows
like $\Theta(n^{7/4})$.

In the recent preprint~\cite{fuchs2024sackin}, analytic combinatorics and the
method of moments are used to fully characterize the asymptotic distribution of
two Sackin indices (using both the length of a longest path from the root and
the length of a shortest path from the root for the depth) in various
combinatorial models of random galled trees with $n$ leaves.
The main result is that the Sackin indices of these random galled trees grow like
$n^{3/2}$ and that, after being appropriately rescaled, they converge
to a well-known probability distribution: the Airy distribution.
An alternative proof of this fact, using branching-process techniques, is
given in Section~\ref{secGHLimits} below.

\subsubsection{The $B_2$ index}

The $B_2$ index was initially defined for trees
in~\cite{shao1990tree}, but was subsequently largely overlooked. However, as
pointed out in~\cite{bienvenu2021revisiting}, in addition to being a valuable
tool for studying certain trees -- see also
\cite{khurana2024limits,kersting2024tree} -- the $B_2$ index is particularly
relevant in the context of networks, because its formal definition and
interpretation as a balance index remain the same as for trees. 

\break

The definition of the $B_2$ index is based on the simple idea of letting
water flow out of the root of a network and down its edges, and then
comparing the amount of water received by each leaf: the more uniform
this distribution, the more balanced the network. Formally, the
$B_2$ index of a phylogenetic network $G$ is defined as
\begin{equation}
  B_2(G) \;=\;  -\sum_{\ell \in L} p_\ell \log_2 p_\ell \,,  
\end{equation}
where the sum runs over the leaves of $G$ and $p_\ell$ is the probability that
a random walk that starts from the root of $G$ and follows its edges --
choosing which outgoing edge to follow uniformly at random and independently of
everything else at every step -- ends in leaf~$\ell$.

The asymptotic distribution of the $B_2$ index of uniform leaf-labeled galled
trees was recently studied in \cite{bienvenu2024b2}, using two independent
approaches: analytic combinatorics, and local limits. We come back to
this in Section~\ref{secLocalLim}, where the branching process approach
is outlined.


\subsection{The method of moments} \label{secMoments}

In probability theory, the method of moments -- not to be confused with the
method of moments from statistics, which is used for parameter estimation -- is
one of the most basic methods to study the convergence in distribution of a
sequence of random variables. Nevertheless, this method has proved very
useful to study summary statistics of random phylogenetic networks, in
particular in the context of combinatorial models.
In a nutshell (see \cite[Section~30]{billingsley1995probability} for a proper
exposition), it consists in proving the convergence in distribution of a random
variable $X_n$ by showing that, for all integer $p \geq 1$, $\alpha_p(n) \defas
\Expec{X_n^p}$ converges to a finite limit $\alpha_p$ as $n\to \infty$.
Importantly, this approach only works if the moment sequence
$(\alpha_p)_{p\geq 1}$ corresponds to a probability distribution that is
\emph{determined by its moments} -- meaning that there is only one 
probability distribution with that moment sequence
(this need not be the case: a simple example of a distribution that is
not determined by its moments is the log-normal distribution).
In practice, one either recognizes
$(\alpha_p)_{p\geq 1}$ as the moment sequence of a known probability
distribution or, else, uses a sufficient condition to ensure that it uniquely
determines a probability distribution (such as the fact that associated
exponential generating function has a positive radius of convergence, see
\cite[Theorem~30.1]{billingsley1995probability}).

Of course, the key for using the method of moments is being able to 
compute the limit of the moments -- and for this there is no general recipe.
Diverse techniques have been used in the literature:
In \cite{bienvenu2022combinatorial} and \cite{fuchs2024limit}, couplings
with a Markov chains are used to obtain recurrences for the moments of
the number of various fringe patterns in uniform leaf-labeled RTCNs. As often,
using factorial moments~-- that is,
$\ExpecBrackets{X (X - 1)\cdots(X - p + 1)}$ -- simplifies calculations
when proving convergence to the Poisson distribution; and using
central moments simplifies calculations when proving convergence to the
normal distribution. In \cite{chang2024enumerative}, \cite{chang2024galled} and
\cite{fuchs2022asymptotic}, elaborate counting arguments involving
ad hoc bijections are used to study the number of reticulations in
various combinatorial models -- including uniform leaf-labeled tree-child
networks (which were long thought to be untractable) in
\cite{chang2024enumerative}.
Finally, in \cite{fuchs2024sackin} and \cite{bienvenu2024b2}, the analytics
combinatorics approach of~\cite{flajolet2009analytic} and the combinatorial
specification of~\cite{bouvel2020counting}
are used to study the asymptotic distribution of the Sackin indices
and $B_2$ index of galled trees.

The fact that it is by essence quantitative is both a strength and a weakness
of the method of moments: indeed, on the one hand when it works it is
usually very precise and provides the type of results that are needed in
concrete applications; on the other hand, it can involve cumbersome
calculations, and does not always give a lot of insight on the results
that it proves. In the rest of this document, we present several alternative
approaches:
\begin{itemize}
  \item In the next section, we introduce a powerful and flexible method
    that can be used to prove convergence to the Poisson distribution
    by computing the first two moments only: the Stein--Chen method.
  \item In Section~\ref{secLimits}, we use a fundamentally different
    approach: instead of studying the statistics of a network of size $n$ and
    then taking~$n$ to infinity, we first take $n$ to infinity to obtain
    a limit network; and then study the statistics of that limit network.
    In addition to being typically less computation-intensive, this approach
    provides an explicit construction of the limit distribution.
\end{itemize}

\subsection{The Stein--Chen method} \label{secSteinChen}

The Stein method was introduced in~\cite{stein1972bound} with
the specific goal of providing error bounds for the approximation of certain
sums of dependent variables by the normal distribution. However, it has
proved to be a powerful tool to study convergence to other distributions.
Here, we focus exclusively on the Stein method for Poisson approximation, often
known as the \emph{Stein--Chen method}.
There are many excellent references on the topic: see in particular
the dedicated book~\cite{barbour1992poisson} and Section~4 of the
monograph~\cite{ross2011fundamentals} for a detailed treatment,
where all the results presented here can be found; or
Section~4.4 of the recent book by Roch~\cite{roch2024modern} for a more
concise introduction.

Let $(X_1, \ldots, X_n)$ be a collection of Bernoulli variables, with
$X_i \sim \mathrm{Bernoulli}(p_{n,i})$, and set $W_n = \sum_{i = 1}^n X_i$ and
$\lambda_n = \Expec{W_n} = \sum_{i=1}^n p_{n, i}$. For ease of notation,
from now on we write $p_i$ for $p_{n, i}$, but it is important to
remember that the $p_i$'s are allowed to depend on $n$. Recall that the
\emph{total variation distance} between two integer-valued random variables $U$
and $V$ can be expressed as
\begin{equation}
  d_{\mathrm{TV}}(U, V) \;=\;
  \frac{1}{2}\, \sum_{k \in \Z} \big|\Prob{U = k} - \Prob{V = k}\big|\,, 
\end{equation}
and that for discrete random variables convergence in total variation is
equivalent to convergence in distribution (for general random variables,
convergence in total variation is stronger than convergence in
distribution).
The Stein--Chen method refers to a collection of upper bounds on the total
variation distance between $W_n$ and the Poisson distribution with
parameter~$\lambda_n$, as a function of the dependency structure of the
variables~$X_i$.

Let us start with the independent case.
When the variables $X_i$ are independent, we have the following bound
(recall that $\lambda_n = \sum_{i = 1}^n p_i$):
\begin{equation} \label{eqSteinChenIndependent}
  d_{\mathrm{TV}}\big(W_n, \mathrm{Poisson}(\lambda_n)\big) \;\leq\;
  \min\!\big(1, \tfrac{\!1}{\,\lambda_n}\big) \, \sum_{i = 1}^n p_i^2 \,,
\end{equation}
which, for $\lambda_n > 1$, improves the classic inequality of Le
Cam~\cite{steele1994lecam}.

When the right-hand side of~\eqref{eqSteinChenIndependent} vanishes
as $n\to\infty$, the distribution of $W_n$ becomes arbitrarily close
to $\mathrm{Poisson}(\lambda_n)$ as $n$ increases.
Unless $\lambda_n$ converges to a finite value, we cannot say that
$W_n$ converges to the Poisson distribution with parameter~$\lambda_n$, because
$\mathrm{Poisson}(\lambda_n)$ is changing as $n \to \infty$;
but by the optimal coupling theorem there exists a
sequence $(Z_n)$ such that $Z_n \sim \mathrm{Poisson}(\lambda_n)$ and
$\Prob{W_n \neq Z_n} \to 0$. Moreover, when $\lambda_n\to \infty$, the
fact that $d_{\mathrm{TV}}(W_n, \mathrm{Poisson}(\lambda_n))$ goes to zero entails
the following central limit theorem:
\begin{equation}
  \frac{W_n - \lambda_n}{\sqrt{\lambda_n}} \;\tendsto[d]{n\to\infty}\;
  \mathcal{N}(0, 1)\,, 
\end{equation}
though note that this is less precise than
$d_{\mathrm{TV}}(W_n, \mathrm{Poisson}(\lambda_n))\to 0$.
For examples of application of these results, see the discussion
of the birth-hybridization process with linear hybridization rate in
Box~\ref{boxLinearHybridization}.

Assume now that the variables $X_i$ are not independent, but that for
each~$i$ there exists a set of indices
$N_i \subset \Set{1, \ldots, n}$ such that
$X_i$ is independent of $(X_j : j \notin N_i)$. 
This is sometimes known as the \emph{dissociated case}, and the sets $N_i$ are
referred to as \emph{dependency neighborhoods}.
In that case, the Stein--Chen bound becomes
\begin{equation} \label{eqSteinChenDissociated}
  d_{\mathrm{TV}}\big(W_n, \mathrm{Poisson}(\lambda_n)\big) \;\leq\;
  \min\!\big(1, \tfrac{\!1}{\,\lambda_n}\big) \,
  \sum_{i = 1}^n \Big(p_i^2 +\sum_{\substack{j \in N_i\\ j \neq i}}
  \!\!\big( p_i p_j + \Expec{X_i X_j}\big) \Big)\,.
\end{equation}
As an example of application, consider the birth-hybridization process with
linear hybridization rate stopped upon reaching $n$ lineages, and let
$U_n$ denote the number of hybridizations that are immediately followed by the
branching of the hybrid (that is, before any other lineage has had time to
branch or hybridize).
This quantity can be expressed as $\sum_{i=2}^{n-2} X_i$, where $X_i$ is the
indicator variable of ``the $i$-th event is a hybridization and the
$(i+1)$-th event is the branching of the hybrid lineage that was just created''.
It is straightforward to see that
\begin{equation}
  \Prob{X_i = 1} \;=\; \frac{\theta(1-\theta)}{i+1} \, , 
\end{equation}
where $\theta$ is the hybridization probability.
As a result, the expected value of $U_n$ is $u_n \defas \Expec{U_n} =
\theta(1-\theta) (H_{n-1}- 3/2)$, where $H_n$ denotes the $n$-th harmonic
number, and therefore $u_n \sim \theta(1-\theta) \log n$.
The random variables $X_i$ and $X_{i+1}$ are not independent, because
$X_i = 1$ implies that the $(i+1)$-th event is a branching, and therefore
$X_{i+1} = 0$. However, $X_i$ is independent of ${(X_j : |i-j| > 1)}$. Thus,
we can use the Stein--Chen bound~\eqref{eqSteinChenDissociated} with the
dependency neighborhoods $N_2 = \{2, 3\}$, $N_{n-2} = \{n-3, n-2\}$ and
$N_i = \{i-1, i, i+2\}$ for $i = 3, \ldots, n-3$. After some simplifications,
this yields the following upper bound on
$d_{\mathrm{TV}}(U_n, \mathrm{Poisson}(u_n))$:
\begin{equation} 
  \frac{\theta(1-\theta)}{u_n} \,
   \sum_{i = 2}^{n-2} \Bigg(\frac{1}{(i+1)^2} +
   \frac{\Indic{i\neq 2}}{i(i+1)} + \frac{\Indic{i\neq n-2}}{(i+1)(i+2)}
  \Bigg) \;\sim\; \frac{C}{\log n} \,, 
\end{equation}
for some positive $C$. We conclude that $U_n$ is asymptotically
Poisson-distributed with parameter $u_n$, in the sense explained above.

\break

Finally, let us assume that the random variables $X_i$ are not independent
and do not have small dependency neighborhoods, but that they are monotonely 
related in the following sense:
We say that the Bernoulli variables $(X_1, \ldots, X_n)$ are
\emph{positively related} if, for each $i = 1, \ldots, n$, there exists a vector
$\smash{(X^{(i)}_1, \ldots, X^{(i)}_n)}$ that has the conditional distribution
of $(X_1, \ldots, X_n)$ given $X_i = 1$ and that satisfies
$\smash{X^{(i)}_j \geq X_j}$ for all $j$.
For such variables, we have
\begin{equation} \label{eqSteinChenPositive}
  d_{\mathrm{TV}}\big(W_n, \mathrm{Poisson}(\lambda_n)\big) \;\leq\;
  \min\!\big(1, \tfrac{\!1}{\,\lambda_n}\big) \,
   \Big(\Var{W_n} -\lambda_n + 2\sum_{i = 1}^n p_i^2\Big).
\end{equation}
Conversely, we say that $(X_1, \ldots, X_n)$ are \emph{negatively related}
if for all $i$ there exists a vector $(\smash{X^{(i)}_j} : j \neq i)$ that has
the conditional distribution of $(X_j : j \neq i)$ given $X_i = 1$
and that satisfies $\smash{X^{(i)}_j} \leq X_j$ for all $j \neq i$. In that
case,
\begin{equation} \label{eqSteinChenNegative}
  d_{\mathrm{TV}}\big(W_n, \mathrm{Poisson}(\lambda_n)\big) \;\leq\;
  \min(1, \lambda_n) \,
   \Big(1 - \frac{\Var{W_n}}{\lambda_n}\Big).
\end{equation}
Note that, using \eqref{eqSteinChenPositive} or \eqref{eqSteinChenNegative},
only the first two moments of $W_n$ are needed to show that it is
asymptotically Poissonian (whereas the methods of moments would require us to
examine all moments).
Also note that we only need to know the \emph{existence} of a monotone coupling
between $(X_1, \ldots X_n)$ and $\smash{(X_1^{(i)}, \ldots, X^{(i)}_n)}$,
not to exhibit one such coupling. In particular,
the existence of such couplings is implied by the notion of
\emph{positive} (resp.\ \emph{negative}) \emph{association}, which are
often easier to establish in practice; see
\cite[Section~2.2]{barbour1992poisson} for a detailed discussion.

\enlargethispage{2ex}

\begin{figure}[h]
  \centering
  \captionsetup{width=.95\linewidth}
  \includegraphics[width=0.6\linewidth]{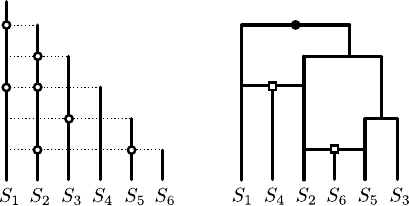}
  \caption{The ``sticks-and-pearls'' construction of birth-hybridization
  networks conditioned to have $n$ leaves. Left, the $n$ sticks together
  with the pearls that have been placed on them as described in the main text.
  Right, the corresponding network, using the same representation as in
  Figure~\ref{figExampleNetwork} and where the leaves are labeled by the
  corresponding stick.}
  \label{figSticksAndPearls}
\end{figure}

As an example of application, consider again the birth-hybridization process
with linear hybridization rate stopped upon reaching $n$ lineages, and
let $R_n$ be the number of reticulation leaves of the corresponding
network, i.e.\ the number of reticulations whose child is a leaf.
Let us adapt the ``sticks-and-pearls'' construction used in
\cite[Section~2.2]{bienvenu2022combinatorial}: let $S_1, \ldots, S_n$ be sticks of
decreasing heights $|S_1| = n, \ldots, |S_n| = 1$ that represent the lineages,
sorted from oldest to newest.  Place a pearl at height $n-1$ on $S_1$. Then,
for each $k = 3, \ldots, n$, independently:
\begin{itemize}
  \item choose a stick $C_k^{(1)}$ uniformly at random from 
    $\{S_1, \ldots, S_{k-1}\}$ and
    place a pearl on it at height $n-k+1$ -- call these pearls
    \emph{unconditional};
  \item with probability $\theta$, choose a second
    stick $C_k^{(2)}$ uniformly at random from 
    $\{S_1, \ldots, S_{k-1}\} \setminus \{C_k^{(1)}\}$ and
    place a pearl on it at height $n-k+1$.
\end{itemize}
Finally, build a network by adding horizontal edges going from each pearl at
height $n-k+1$ to the upper tip of $S_k$, as illustrated in
Figure~\ref{figSticksAndPearls}.

Let $I_i$ be equal to 1 if the parent of the leaf corresponding to $S_i$ is
a reticulation, and 0 otherwise -- so that $R_n = \sum_{i = 1}^n I_i$.
Note that $I_1 = I_2 = 0$ and that, for each $i \geq 3$, $I_i = 1$ if
and only if (1) two pearls were placed at height $n-i+1$ and (2) at each
step $k = i+1, \ldots, n$ of the construction, no pearl was placed on~$S_i$.
The dependency between the variables $I_i$ is complex: on the one hand,
$I_i = 1$ implies that two pearls were placed at height $n-i+1$, and that
the stick $S_i$ was never chosen at steps $k > i$ of the procedure
-- which tends to increase the number of pearls falling on
other sticks, and thus to decrease the probability that $I_j = 1$ for $j \neq i$.
On the other hand, $I_i = 1$ decreases the probability that two pearls
(as opposed to one) were placed at steps $k > i$ -- which tends
to increase the probability that $I_j = 1$ for $j \neq i$. Nevertheless,
it is still possible to use the Stein--Chen method to study $R_n$.

Write $A_i$ for the event that two pearls were placed at height $n-i+1$ and
that no unconditional pearl was placed on $S_i$ at steps $k = i+1, \ldots, n$
of the construction; and write $B_i$ for the event that no
reticulation-inducing pearl was placed on $S_i$ at steps $k = i+1, \ldots, n$.
Thus, $I_i$ is the indicator variable of $A_i\cap B_i$.
Let $X_i$ be the indicator variable of $A_i$. It is straightforward to see
that
\begin{equation} \label{eqProbXiEq1}
  \Prob{X_i = 1} \;=\;
  \theta \,\times\! \!\prod_{k = i+1}^{n} \frac{k-2}{k-1}
  \;=\; \theta\, \frac{i-1}{n-1} \,.
\end{equation}
Similarly, for $i < j$ we have
\begin{align} \label{eqProbXiXjEq1}
  \Prob{X_i = 1, X_j = 1} 
  \;&=\;\;
  \theta^2 \times \Bigg[\prod_{k = i+1}^{j} \frac{k-2}{k-1} \Bigg] \times 
  \Bigg[\prod_{k = j+1}^{n} 
    \frac{k-3}{k-1}\Bigg]\nonumber \\
  &=\;\;\theta^2 \frac{(i-1)(j-2)}{(n-2)(n-1)} \, .
\end{align}
With these expressions, the mean and variance of
$W_n \defas \sum_{i = 3}^n X_i$ can be computed.
Moreover, the random variables
$(X_3, \ldots, X_n)$ are negatively related (as mentioned above,
conditioning on a stick not receiving pearls increases the probability
that other sticks receive pearls; see the \SuppMat{} for
details).
As a result, Equation~$\eqref{eqSteinChenNegative}$ can
be used to bound the total variation distance between $W_n$
and the Poisson distribution with parameter $\lambda_n \defas \Expec{W_n}$.
For fixed $\theta$,
we have $\lambda_n \sim \theta n / 2$ as $n\to \infty$ and
$\Var{W_n}\sim \theta (6 - 5\theta) n / 12$, and
the right-hand side of~\eqref{eqSteinChenNegative}
does not go to 0. However, if $\theta = \theta_n$ is allowed
to depend on $n$ in such a way that $\theta_n \to 0$ as $n \to \infty$,
then $\Var{W_n} \sim \lambda_n$ and the total variation distance between
$W_n$ and $\mathrm{Poisson}(\lambda_n)$ goes to zero, 
showing that for small values of~$\theta$ and large values of~$n$
the distribution of $W_n$ is well-approximated by the Poisson distribution with
parameter~$\lambda_n$.
In particular, for $\theta_n = c/n$ we have 
$W_n \to \mathrm{Poisson}(c)$ in distribution as $n\to\infty$.

Finally, let $Y_i$ be the indicator variable of $A_i \setminus B_i$,
and note that $I_i = X_i - Y_i$, so that
$R_n = W_n - \Delta_n$ where $\Delta_n = \sum_{i = 3}^n Y_i$.  Since
$\Expec{Y_i} \leq \theta^2$, we have
\[
  \Prob{\Delta_n > 0} \;\leq\; \Expec{\Delta_n} \;\leq\; \theta^2 n,
\]
which entails that $\Delta_n$ converges to 0 in probability when
$\theta^2 n \to 0$. In particular, for $\theta_n = c/n$, by
Slutsky's theorem we get that the number $R_n$ of reticulation leaves
converges in distribution to $\mathrm{Poisson}(c)$
as $n\to\infty$.

\section{Blowups of Galton--Watson trees} \label{secBlowups}

In this section, we focus on a class of models of random phylogenetic networks
that will play a crucial role in Section~\ref{secLimits}:
blowups of Galton--Watson trees. The first use of these models to study
phylogenetic networks is due to Stufler in \cite{stufler2022branching}.
They have since been used in \cite{bienvenu2024zero} to study pattern
occurrences in phylogenetic trees and networks; in \cite{bienvenu2024b2}
to study the $B_2$ index of galled trees; and in \cite{bienvenu2024branching}
to study a model of random phylogenetic network generated by a
birth-death-coalescence process with mutation.
As can been seen from the variety of models treated in these examples,
blowups of Galton--Watson trees are a unifying framework that can
be used in very diverse settings. 

\subsection{Size-conditioned Galton--Watson trees} \label{secDefBlowup}

The goal of this section is to recall some classic definitions and to set
some notation.
Let us start with the definition of Galton--Watson trees.
Let $\xi$ be an integer-valued random variable, and to avoid trivialities
assume that $\Prob{\xi = 0} > 0$
and \mbox{$\Prob{\xi = 1} \neq 1$}.

A \emph{Galton--Watson tree with offspring distribution $\xi$} is a
random rooted tree obtained as follows:
start with one vertex in generation~$0$ (the root); then, for 
$t = 0, 1, 2\ldots$ and for each vertex~$v$ in generation~$t$, let $v$ have
$\xi_v$ children in generation~$t+1$, where $\xi_v$ is an independent
realization of $\xi$.
Note that we view this process as generating \emph{ordered trees},
i.e.\ trees where the children of each vertex $v$ are labeled from 1 
to~$\xi_v$. It immediately follows from the definition of a Galton--Watson
tree $T$ with offspring distribution $\xi$ that, for any fixed ordered
tree~$\mathbf{t}$, 
\begin{equation} \label{eqProbaGW}
  \Prob{T = \mathbf{t}} \;=\; \prod_{v \in \mathbf{t}} \Prob{\xi = d^+{(v)}} \,,
\end{equation}
where the product runs over all vertices of $\mathbf{t}$ and $d^+(v)$
denotes the outdegree of $v$ in~$\mathbf{t}$.
We write $T\sim\mathrm{GW}(\xi)$ to indicate that $T$ is a Galton--Watson
tree with offspring distribution~$\xi$.

A Galton--Watson tree may or may not be finite.
With our assumptions on~$\xi$, we have the following phase-transition in
that respect:
a Galton--Watson tree with offspring
distribution $\xi$ is finite
with probability~1 if $\Expec{\xi} \leq 1$, and infinite with positive
probability if $\Expec{\xi} > 1$. Consequently, a tree $T\sim \mathrm{GW}(\xi)$
with $\Expec{\xi}=1$ is known as a \emph{critical Galton--Watson tree}.
These trees will play a central role in Section~\ref{secLimits}.

Depending on the context, the \emph{size} of a tree $\mathbf{t}$ could refer to its
number of vertices or to its number of leaves; for now, we leave this
unspecified and we simply denote it by~$|\mathbf{t}|$. Provided that
$\Prob{|T| = n} > 0$, a Galton--Watson tree $T$ can be conditioned to have
size~$n$: the resulting tree $T_n$ is called a \emph{size-conditioned
Galton--Watson tree}, and we write $T_n \sim \mathrm{GW}_n(\xi)$. If $T$ is a
Galton--Watson tree, we always denote the corresponding size-conditioned tree
by $T_n$.

Size-conditioned Galton--Watson trees are an instance of a more general
class of random trees known as \emph{simply generated trees}:
let $\mathbf{w} = (w_i)_{i\geq 0}$ be a sequence of nonnegative numbers 
known as \emph{weights}, and assume that
$w_0 > 0$ and that there exists $k$ such that $w_k > 0$.
Define the weight of a finite tree~$\mathbf{t}$ to
be
\begin{equation}
  w(\mathbf{t}) \;=\; \prod_{v\in \mathbf{t}} w_{d^+(v)} \,.
\end{equation}
Finally, let $\mathcal{T}_n$ denote the set of trees of size~$n$. A simply
generated tree with weight sequence $\mathbf{w}$ is a random tree
$T$ sampled from $\mathcal{T}_n$ with probability proportional to its weight --
that is,
\begin{equation} \label{eqSimplyGenerated}
  \forall \mathbf{t} \in \mathcal{T}_n, \quad
  \Prob{T = \mathbf{t}} \;=\;
  \frac{w(\mathbf{t})}{\sum_{\mathbf{t'} \in \mathcal{T}_n} w(\mathbf{t'})}\,.
\end{equation}
Comparing this with Equation~\eqref{eqProbaGW}, we see that a
size-conditioned Galton--Watson tree with offspring distribution $\xi$ is
a simply generated tree with weight sequence $w_i = \Prob{\xi = i}$.
Conversely, a simply generated trees whose weight sequence defines a probability
distribution is a size-conditioned Galton--Watson tree.

It is possible to modify a weight sequence without changing the
distribution of the corresponding simply generated tree, through an operation
known as \emph{exponential tilting}.
Indeed, it is straightforward to check from
Equation~\eqref{eqSimplyGenerated} that if the size of a tree is defined
as its number of vertices, then any weight sequence $\tilde{\mathbf{w}}$ given
by $\tilde{w}_i = C \zeta^i w_i$ for positive constants
$C$ and $\zeta$ defines the same simply generated tree
as $\mathbf{w}$. Similarly, if the size is defined as the number of leaves,
then using the modified weights $\hat{w}_0 = C$ and
$\hat{w}_i = \zeta^{i-1} w_i$ for $i\geq 1$ does not change the tree. As a
result, many (though not all) simply generated trees can also be expressed as
size-conditioned Galton--Watson trees.

Many classic combinatorial models of random trees are (or can be obtained from)
simply generated trees\,/\,size-conditioned Galton--Watson trees,
see~\cite[Section~10]{janson2012simply} for a list of examples. In particular,
this includes both the rooted and unrooted leaf-labeled binary trees often
encountered in mathematical phylogenetics; see
e.g.~\cite[Proposition~8]{bienvenu2024zero} for a detailed proof of this
well-known fact.

\subsection{Blowups of trees} \label{secBlowupGWT}

Here we present the general setting of blowup constructions, with an
emphasis on applications to phylogenetic networks. See the monograph
\cite{stufler2020limits} for a detailed treatment in a more general setting. 

\begin{figure}[h]
  \centering
  \captionsetup{width=.95\linewidth}
    \includegraphics[width=0.95\linewidth]{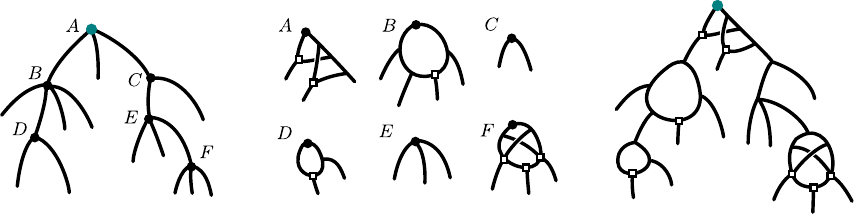}
  \caption{Illustration of the blowup procedure.
  Left, the base tree $T$. Middle,
  the decorations, i.e.\ the phylogenetic networks (here, not necessarily binary,
  see~$E$ and~$F$) associated to the internal vertices of $T$. Right, the
  phylogenetic network obtained by replacing the vertices of $T$ by their
  decoration.  Compare with Figure~\ref{figGalledTree}.}
  \label{figBlowup}
\end{figure}

A \emph{blowup of a random tree $\,T$} is any random phylogenetic network
obtained by, first, sampling $T$; and then 
replacing each internal vertex of $T$ by a random phylogenetic network $N_v$
with $d^+(v)$ leaves, whose distribution depends on $d^+(v)$ only and that
is sampled independently of everything else. To perform this ``replacement''
of $v$ by $N_v$, the root of $N_v$ is identified with $v$ and the
leaves of $N_v$ are identified with the children of~$v$. In doing so, the
leaves of $N_v$ are matched to the children of $v$ uniformly at random.
See Figure~\ref{figBlowup} for an illustration.

The tree from which a blowup is built is called its \emph{base tree},  and
the random networks associated to the vertices of the base tree are called
the \emph{decorations} of the vertices. Thus, a blowup is parametrized by
a random tree $T$ and a family $\nu = (\nu_k)_{k\geq 1}$ 
such that each $\nu_k$ is a probability distribution on the set of 
finite phylogenetic networks with $k$ leaves.

Several models of random phylogenetic networks can be obtained as blowups of
simply generated trees. Stufler proved in~\cite{stufler2022branching} that this
includes uniform leaf-labeled level-$k$ networks, and that size-conditioned
critical Galton--Watson trees can be used for the base tree, which makes it
possible to access a wide range of probabilistic tools.
More generally, any model of random phylogenetic network that
corresponds to the uniform distribution on a class of leaf-labeled
phylogenetic networks defined through restrictions on their blocks only
can be obtained as the blowup of a simply generated tree.

More specifically, recall the notions of
biconnected network and of block from Section~\ref{secGalledTrees}, and
let $\mathcal{B}$ be a class of biconnected leaf-labeled
phylogenetic networks such that the set $\mathcal{B}_n$ of elements of
$\mathcal{B}$ with $n$ leaves is finite for all $n$.
Let $\mathcal{A}$ be the class of
leaf-labeled phylogenetic networks whose blocks belong to $\mathcal{B}$.
Then, the uniform distribution on the set $\mathcal{A}_n$ of elements of 
$\mathcal{A}$ with $n$ leaves can be obtained as the blowup of the
simply generated tree with weight sequence $w_0 = 1$ and
$w_i = |\mathcal{B}_i| / i!$, using the uniform distribution on
$\mathcal{B}_i$ to sample the decorations.
Whether this simply generated tree is also a size-conditioned
Galton--Watson tree depends on whether there exists an exponential tilting of
the weight sequence that turns it into a probability distribution,
i.e.\ on the asymptotic growth of the sequence $|\mathcal{B}_i|$.
When it is possible to get a probability distribution with mean~1,
the model is guaranteed to be at least somewhat tractable, as techniques
presented in Section~\ref{secEnumAndSampling}, \ref{secLocalLim} and
\ref{secGHLimits} are guaranteed to apply -- in particular (but not only)
in the case where the offspring distribution has a finite variance.
The level of detail to which we can understand the model will then depend
on our understanding of the combinatorial class $\mathcal{B}$.
For instance, in the case of galled trees $\mathcal{B}$ is the class of galls
and the construction is fully explicit; see~\cite[Section~3.4]{bienvenu2024b2}.

Note that the use of blowup constructions is not limited to combinatorial models
and situations where the decorations are chosen from a finite
set of biconnected components: they can also be useful to study process-based
models in situations where $\mathcal{B}_i$ is infinite.
In that case, one typically needs to show that the distance from
a leaf chosen uniformly at random in the network
to the root of the corresponding decoration
has finite exponential moments -- see
\cite{bienvenu2024branching}.

\subsection{Enumeration and sampling} \label{secEnumAndSampling}

We now briefly mention two consequences of the fact that
a combinatorial model admits a construction as a blowup of a critical
Galton--Watson tree; other consequences concerning the geometry of the
corresponding networks are discussed in more detail in the next section. 

The first consequence concerns the asymptotic enumeration of the corresponding
combinatorial class. In a nutshell, if a class $\mathcal{A} = \bigcup_n
\mathcal{A}_n$ of phylogenetic networks can be obtained as a blowup of a
size-conditioned critical Galton--Watson tree whose offspring distribution
$\xi$ has a finite variance, the asymptotics of its counting sequence $a_n
\defas |\mathcal{A}_n|$ are governed by its tree-like structure and are of the
form
\begin{equation}
  a_n  \;\sim\; c\, n^{-\frac{3}{2}} \rho^{-n}\, n! \,, 
\end{equation}
with the presence of the $n^{-3/2}$ subexponential factor that is
characteristic of trees,
and an exponential bias by $\rho = \zeta\,\Prob{\xi = 0}$, where
$\zeta$ is the parameter of the exponential tilting used to get a critical
probability distribution from the weight sequence.
In slightly more detail, this result comes from the observation that the
exponential generating function $A(z) = \sum_{n\geq 0} a_n z^n / n!$
satisfies
\begin{equation} \label{eqGeneratingFunctionA}
  A(z) \;=\; z \;+\; B(A(z)) \,, 
\end{equation}
where $B$  is the exponential generating function of the combinatorial
class $\mathcal{B}$ through which $\mathcal{A}$ is defined; and that the
probability generating function $G(z) = \Expec*{\normalsize}{z^{|T|}}$ of
the number of leaves of a tree $T\sim\mathrm{GW}(\xi)$ satisfies
\begin{equation} \label{eqGeneratingFunctionG}
  G(z) \;=\; z \, \Prob{\xi = 0} \;+\; \zeta^{-1} B(\zeta G(z))\,.
\end{equation}
Together, Equations~\eqref{eqGeneratingFunctionA}
and~\eqref{eqGeneratingFunctionG} imply $A(z) = \zeta G(z/\rho)$. The claimed
result then follows from the standard fact that, since we have assumed that
$\xi$ has a finite second moment, the number of leaves of $T$ satisfies
$\Prob{|T| = n} \sim K n^{-3/2}$, where the constant $K$ is
explicit~\cite[Theorem~3.1]{kortchemski2012invariance}.
See~\cite[Lemma~1.1]{stufler2022branching} for a detailed derivation in the
case of level-$k$ networks that is easily generalized to the framework
considered in this section.

The second consequence of the existence of a blowup construction -- which we
only mention in passing -- is that,  provided the decorations can be sampled
efficiently, it is possible to sample the corresponding networks in expected
linear time. See~\cite{panagiotou2023exact}, in particular Section~5.2.

\section{Capturing the geometry of large networks} \label{secLimits}

We now present powerful tools to study the geometry of large
networks: local limits and Gromov--Hausdorff scaling limits. In both
cases, the motivation is the same: to define a limit network from which
the asymptotics of statistics of interests can be read directly.
In other words, suppose we are interested in some summary statistic $f$ of a
random phylogenetic network $G_n$ of size $n$, and want to study the asymptotic
behavior of $f(G_n)$ as $n$ becomes large. Instead of studying
$f(G_n)$ for finite $n$ and then taking the limit as $n$ goes to infinity,
we first identify the limit network $G_\infty$ and then compute $f(G_\infty)$.
When it works, this approach can have several advantages:
\begin{itemize}
  \item when studying several statistics, it
    can be more efficient, because some of the reasoning and calculations
    get ``factored-in'' in the determination of the limit network $G_\infty$;
  \item it is often more robust:  indeed, studying summary statistics, e.g,
    through the method of moments often requires ad hoc ``tricks'' that
    sometimes depend on fine details of the model and\,/\,or of the statistic of
    interest, and may not be applicable for related models\,/\,statistics;
  \item it is sometimes easier, and typically less computation-intensive --
    especially in situations where the finite networks have a dependency
    structure that vanishes in the limit.
    Size-conditioned Galton--Watson trees and their blowups are a prime
    example of this, as we will see below.
\end{itemize}

Of course, this approach has its own challenges. First, one must be able to
establish convergence to a tractable limit. This is usually straightforward
for blowups of size-conditioned Galton--Watson trees, as these have
universal limits, under minimal assumptions. Second, the approach requires
continuously extending the summary statistics of interest to the limit space, 
which may not be possible. However, the fact that the extension of a
statistic $f$ is not continuous does not necessarily doom the approach,
as it may be continuous along sequences of interest; the $B_2$ index provides
an example of this.

Finally, the notions of local limit and Gromov--Hausdorff limit are
complementary:
the local limit captures the structure of networks on the small scale,
around a focal point of interest, and is suited to study statistics
such as average degrees or pattern occurrences; 
whereas the Gromov--Hausdorff limit captures the large-scale geometry
and is suited to study statistics such as the diameter or typical distances
between points.

\subsection{Small-scale geometry: local limits} \label{secLocalLim}

The notion of local limit -- also known as the
\emph{Benjamini--Schramm limit} -- was introduced by Benjamini and
Schramm in \cite{benjamini2001recurrence} and independently reintroduced
by Aldous and Steele in \cite{aldous2004objective}. It has since become a
standard tool in probability theory. We refer the reader to
\cite[Chapter~2]{vanderhofstad2024random} for a general introduction,
and to \cite{janson2012simply} and~\cite{abraham2015introduction} for
a detailed treatment in the context of Galton--Watson trees.

The idea is to study the convergence of the structure of networks
around finite neighborhoods of their root. More specifically, for any phylogenetic
network $\mathbf{g}$, let $[\mathbf{g}]_k$ denote the subgraph of $\mathbf{g}$
induced by the vertices at distance at most $k$ from the root.
We say that a sequence of phylogenetic networks $(\mathbf{g}_n)$ \emph{converges locally},
to the (possibly infinite) phylogenetic network $\mathbf{g}_\infty$ if, for all
$k\geq 0$, there exists $N\geq 0$ such that
$[\mathbf{g}_n]_k = [\mathbf{g}_\infty]_k$ for all $n \geq N$.
This defines a well-behaved notion of convergence on the set of phylogenetic
networks and makes it possible to define a corresponding notion of
the convergence in distribution
for the local topology~-- also known as $\emph{local weak convergence}$ -- of a
sequence of random phylogenetic networks.
Note that here we have defined local convergence around the root of the
network, but that it is also possible to define it around other focal points of
interest (such as a vertex sampled uniformly at random among vertices\,/\,the
leaves).

\begin{figure}[h]
  \centering
  \captionsetup{width=.95\linewidth}
  \includegraphics[width=0.5\linewidth]{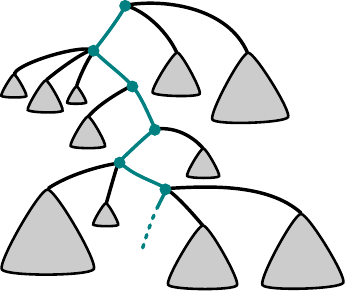}
  \caption{The structure of Kesten's tree $T_*$. The spine of the tree is highlighted
  in green, and each grey triangle represents an independent copy of
  the critical Galton--Watson tree~$T$ from which $T^*$ is built. All these
  trees are finite with probability~1.
  Along the spine, the offspring distribution is the size-biased offspring
  distribution of~$T$.}
  \label{figKesten}
\end{figure}

Size-conditioned critical Galton--Watson trees have a universal local limit
known as \emph{Kesten's tree}. To describe this tree, recall that
the \emph{size-biased distribution} of a nonnegative integer-valued random
variable $\xi$ is the probability distribution $(\hat{p}_k)_{k\geq 1}$ defined
on the positive integers by
$\hat{p}_k = k\,\Prob{\xi = k} \Expec{\xi}^{-1}$. We write $\hat{\xi}$
to denote a random variable with the size-biased distribution of $\xi$.
The Kesten tree associated to $T\sim\mathrm{GW}(\xi)$
is the infinite random tree $T_*$ generated by the following
two-type Galton--Watson process: vertices are either of type \emph{spine} or of
type \emph{regular}. Starting with one spine vertex in generation~0, for all
$t = 0, 1, 2, \ldots$ let each vertex $v$ in generation~$t$ have:
\begin{itemize}
  \item $\xi_v$ regular children, if $v$ is a regular vertex;
  \item one spine child and $\hat{\xi}_v - 1$ regular children, if $v$ is a
    spine vertex,
\end{itemize}
where each $\xi_v$ (resp.\ $\hat{\xi}_v$) is a copy of $\xi$ (resp.\
$\hat{\xi}$) that is independent of everything else 
(and where the position of each spine vertex is chosen uniformly at random
among the children of its parents, to get an ordered tree).
Note that the spine vertices form an infinite path known as the \emph{spine}
of~$T_*$. See Figure~\ref{figKesten}.

The fundamental result on the local convergence of Galton--Watson trees is
that if $T$ is a critical Galton--Watson tree and $T_n$ is the corresponding
size-conditioned Galton--Watson tree -- where the size can be the number of
vertices or the number of leaves
(or more complex quantities, see~\cite{abraham2015introduction}) -- then
$T_n$ converges in distribution to the Kesten tree $T_*$ associated to $T$.

The local convergence of Galton--Watson trees directly implies the local
convergence of their blowups: more generally, if $T_n$ is a
random tree (not necessarily a conditioned Galton--Watson tree) that converges
to some local limit $T_\infty$ and if $G_n$ is a blowup of $T_n$ with respect
to a family of decorations $\nu$, then $G_n$ converges locally to the blowup of
$T_\infty$ with respect to~$\nu$ -- see \cite{bienvenu2024b2}.

Statistics that depend on a finite neighborhood of the root
(such as its degree) are continuous for the local topology. However,
as mentioned at the beginning of this section, many summary statistics
used in phylogenetics are not continuous for the local topology. These
include pattern counts: indeed, patterns
that are present in every network of a sequence $(G_n)$ can
``escape to infinity'' as the networks become large, and thus be invisible for
the local limit. As a simple example of this, let $B_h$ denote the complete
binary tree with height $h$ and $2^{h}$ leaves. The local limit of $B_h$
as $h \to \infty$ is the infinite binary tree, which has no leaves.
As a result, the number of leaves is not continuous for the local topology.

Nevertheless, statistics that are not continuous for the local topology
can sometimes be studied using local limits. For instance,
in~\cite{bienvenu2024zero} local limits are used to study
pattern occurrences in phylogenetic trees and networks; and in
\cite{bienvenu2024b2} they are used to study the $B_2$ index of blowups of
Galton--Watson trees (in particular, if
$T_n \sim \mathrm{GW}_n(\xi)$, where $\xi$ is critical and has a finite third
moment, then
the $B_2$ index of a blowup of $T_n$ converges in distribution and in all moments
to the $B_2$ index of the blowup of the corresponding Kesten tree $T_*$ --
see \cite[Theorem~3.7]{bienvenu2024b2}).
A criterion that can be used to check whether the asymptotic behavior of a summary
statistic is determined by the local weak limit is given
in~\cite{kurauskas2022local}

As a simple example of application showing how local limits can be used
to simplify calculations, consider the $B_2$ index of a tree
$T_n$ generated by the PDA model, i.e.\ that is uniformly distributed
on the set of rooted binary trees with $n$ labeled leaves. Using direct
but somewhat tedious calculations that require some guesswork,
it is possible to show that
$\Expec{B_2(T_n)} = 3(n-1)/(n+1)$, see
\cite[Theorem~3.9]{bienvenu2021revisiting}.
Now, let us use the local limit to prove that
$\ExpecBrackets{B_2(T_n)} \to 3$ as $n\to\infty$.
First, recall the well-known fact that $T_n$ can be obtained as a
size-conditioned Galton--Watson tree with offspring distribution $\xi =
2\times\mathrm{Bernoulli}(1/2)$, see e.g.~\cite{lambert2017probabilistic}.
Note that the size-biased distribution of $\xi$ is deterministic equal
to~2.  As a result, the local limit $T_*$ of $T_n$ has a simple recursive
structure: it consists of a root vertex with two children, one of which is the
root of an independent copy of $T_*$ and the other of an independent
$\mathrm{GW}(\xi)$ tree $T$.
Thus, by using the grafting property of the $B_2$ index
\cite[Corollary 1.12]{bienvenu2021revisiting} we get
the following distributional equation for $B_2(T_*)$:
\begin{equation}\label{eqExampleB201}
  B_2(T_*) \;\overset{d}{=}\;\frac{1}{2}\big(B_2(T_*) \;+\;B_2(T)\big) \;+\; 1\,,
\end{equation}
where $T_*$ and $T$ are independent on the right-hand side. Similarly,
using that $T$ is reduced to its root with probability $1/2$ and obtained by
grafting two independent copies of itself on the root otherwise, we get
\begin{equation}\label{eqExampleB202}
  B_2(T) \;\overset{d}{=}\;
  \Big( \frac{1}{2} \big(B_2(T) \;+\; B_2(T')\big) \;+\; 1 \Big) \, X\,,
\end{equation}
where $X\sim \mathrm{Bernoulli}(1/2)$, $T'$ is a copy of $T$, and $\{T, T', X\}$
are independent. Taking expectations, we get
$\ExpecBrackets{B_2(T)} = \frac{1}{2}(\ExpecBrackets{B_2(T)} + 1)$, i.e.\
$\ExpecBrackets{B_2(T)} = 1$;
plugging this in Equation~\eqref{eqExampleB201}
yields $\ExpecBrackets{B_2(T_*)} = 3$.
The same approach is used in~\cite{bienvenu2024b2} to study the $B_2$ index
of galled trees.

\subsection{Large-scale geometry: Gromov--Hausdorff limits} \label{secGHLimits}

In contrast to local limits, which describe the structure of networks around
their root on its original scale,
Gromov--Hausdorff limits are \emph{scaling limits}, meaning that they
describe the structure of networks as we see them after ``zooming out''
until they fit in our field of view; in doing so, we lose the fine
details but get a better view of the global structure.
This approach requires encoding the structure of networks in a way that makes
it possible to perform this scaling.
To that end, we view networks as metric spaces. In the case of
combinatorial models, we simply
consider the usual graph distance on the corresponding undirected graph; in
the case of process-based models, it is often relevant to
give lengths to the edges that match the time-embedding of the network, as
in the definition of temporal networks from Section~\ref{secVocab}.

The definition of the Gromov--Hausdorff limit is technical, so we only
sketch the main idea here.
We refer the reader to \cite[Chapter~4]{evans2008probability} or
\cite[Section~6]{miermont2009tessellations} for a detailed introduction.
Loosely speaking, we say that a sequence of compact metric spaces $(\mathcal{M}_n)$
converges to the compact metric space $\mathcal{M}$ in the Gromov--Hausdorff
sense if $\mathcal{M}_n$ and $\mathcal{M}$ can be embedded in a common metric
space in such a way that they overlap arbitrarily well as $n$ goes to infinity.
Slightly more formally, recall that an isometry is a distance-preserving
bijection between two metric spaces, which are then said to be isometric. For
our purposes, two isometric spaces can be considered identical; we
therefore identify metric spaces with their isometry class.
The preliminary to the notion of Gromov--Hausdorff convergence is the
\emph{Hausdorff distance} between two compact sets $A$ and $B$
of a metric space $\mathcal{E} = (E, d)$. It is defined as
\begin{equation}
  d_{\mathrm{H}}(A, B) \;=\;\max\Set{\sup_{a\in A} d(a, B),\; \sup_{b\in A} d(b, A)}\,, 
\end{equation}
where $d(a, B) = \inf_{b\in B} d(a, b)$. The \emph{Gromov--Hausdorff distance}
between two compact metric spaces $\mathcal{M}$ and $\mathcal{M'}$ is then
defined as the infimum of $d_{\mathrm{H}}(M, M')$ 
over all metric spaces $\mathcal{E} = (E, d)$ and all
isometric embeddings of $\mathcal{M}$ and $\mathcal{M'}$ in 
$\mathcal{E}$ as $M\subset E$ for $\mathcal{M}$ and 
$M'\subset E$ for $\mathcal{M}'$.

Due to the abstract nature of its definition, the Gromov--Hausdorff
distance is in general non-explicit: its main purpose is
simply to define the notion of Gromov--Hausdorff convergence 
 -- which is then proved using upper bounds
on the Gromov--Hausdorff distance, rather than by computing it.
Gromov--Hausdorff convergence is a strong notion of convergence and
implies the convergence of essentially all statistics that describe the
large-scale structure of a metric space.

\begin{figure}[h]
  \centering
  \captionsetup{width=.95\linewidth}
  \includegraphics[width=0.95\linewidth]{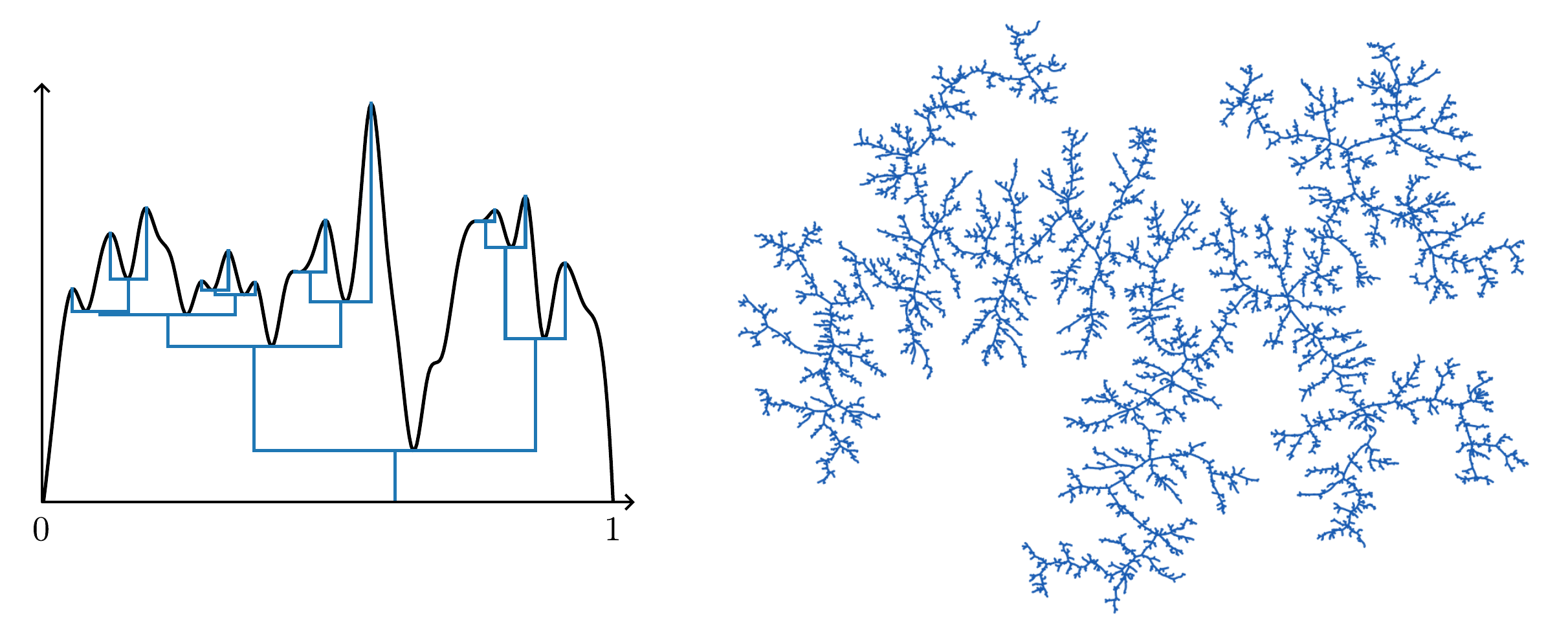}
  \caption{Left, a continuous function $f: [0, 1]\to \R_+$ such that
  $f(0) = f(1) = 0$, and the corresponding $\R$-tree in blue (i.e.\ the metric
  space associated to the (pseudo)distance on $[0, 1]$ defined by
  $d_f(x, y) = f(x) + f(y) - 2\inf_{z \in [x, y]} f(z)$; note that
  horizontal edges have length~0).
  When $f$ is a standard Brownian excursion, this $\R$-tree is the CRT.
  Right, the CRT, as approximated by a critical Galton--Watson tree conditioned
  to have 40000 vertices. Computer code courtesy of Jean-Jil Duchamps.}
  \label{figCRT}
\end{figure}

Size-conditioned critical Galton--Watson trees have universal
Gromov--Hausdorff limits known as \emph{$\alpha$-stable trees}.
In the case where the offspring distribution has a finite variance, 
this limit is the \emph{Brownian continuum random tree} (CRT) --
a random metric space introduced by Aldous in a series of seminal
papers~\cite{aldous1991continuumI, aldous1991continuumII, aldous1993continuum}
that has since become a cornerstone of Brownian geometry. The simplest
way to define it is to take a standard Brownian excursion
$(\mathbf{e}(t))_{t\in[0, 1]}$ and to consider the metric space
induced on $[0, 1]$ by $d_{\mathbf{e}}(x, y) \defas \mathbf{e}(x) + \mathbf{e}(y) -
2\inf_{z \in [x, y]} \mathbf{e}(z)$, identifying points
at distance~0 to get a proper metric.
The resulting metric space is a special type of metric space
known as a \emph{$\R$-tree}, and has a fractal geometry;
see Figure~\ref{figCRT}.

The fundamental result on scaling limits of Galton--Watson trees is that
if $T_n$ is a size-conditioned critical Galton--Watson tree whose offspring
distribution has finite variance $\sigma^2$, the corresponding metric space,
rescaled by multiplying all distances by
$\tfrac{1}{2} \sigma n^{-1/2}$, converges to the CRT
for the Gromov--Hausdorff topology
[\citenum{aldous1993continuum}; see also \citenum{duquesne2003limit,
kortchemski2012invariance}].
Unlike local convergence, the Gromov--Hausdorff convergence
of a tree does not imply the convergence of its blowup: indeed, one has to
check that the distances are not too distorted by the blowup procedure.
Proposition~3.4 and Lemma~3.10 in \cite{bienvenu2024branching} provide a
general recipe for doing so.
In a nutshell, a sufficient condition to ensure the convergence to the CRT
of blowups of $\mathrm{GW}_n(\xi)$ trees with $\Var{\xi} <\infty$
is that the height of a random decoration with $\smash{\hat{\xi}}$ leaves
should have finite exponential moments.

\break

As an example of application of scaling limits to phylogenetics, let us
briefly explain how the Gromov-Hausdorff convergence of
uniform leaf-labeled level-$k$ networks to the CRT, which was proved by 
Stufler in~\cite{stufler2022branching}, can be used to recover a recent 
result concerning the Sackin indices of galled trees (see also
Remark~5 in \cite{fuchs2024sackin}).

In \cite{fuchs2024sackin}, Fuchs and Gittenberger used the method of
moments to prove -- among other things -- that if we consider a
uniform galled tree with $n$ labeled leaves, then letting
$S^{\min}_n$ and $S^{\max}_n$ denote the Sackin indices of that network
(obtained by using the lengths of shortest vs longest paths 
for the depths, respectively -- see Section~\ref{secSackin}), we have
\begin{equation} \label{eqCVSackinIndexGalledTrees}
  \frac{S^{\square}_n}{\sqrt{\pi} \mu^{\square } n^{3/2}}
  \;\tendsto[d]{n\to\infty}\; A\,,\qquad \square \in \{\min, \max\},
\end{equation}
where the random variable $A$ has the Airy distribution and the constants
$\mu^{\min}$ and $\mu^{\max}$ are explicit.
Now, let $\mathbf{g}$ be a galled tree with $n$ leaves $\ell_1, \ldots, \ell_n$,
in depth-first order.
For $\square \in \{\min, \max\}$, let $h^{\square}_{\mathbf{g}}(\ell_i)$ denote the
corresponding notion of depth of~$\ell_i$ -- also known in that context as its
\emph{height}. Following \cite{stufler2022branching}, let the \emph{height
process formed using the leaves} of $\mathbf{g}$ be the piecewise constant
interpolation of the vector $(h^{\square}_{\mathbf{g}}(\ell_i): i = 1,
\ldots, n)$, seen as a function on $\COInterval{0, n}$ -- that is,
\begin{equation}
  h^\square_{\mathbf{g}}(t) \;\defas\;
  \sum_{i = 1}^{n} h^{\square}_{\mathbf{g}}(\ell_i) \, \Indic*{\COInterval{i-1, i}}(t)\,,  
  \qquad t\in \COInterval{0, n}
\end{equation}
(note that, to avoid some technicalities, linear interpolation rather than
piecewise-constant interpolation is used in \cite{stufler2022branching};
this is irrelevant because these two versions of the height process have the
same scaling limit). With this definition, the Sackin index of $\mathbf{g}$
is exactly the area under the curve of its height process. 
For {${\square = \min}$}, i.e.\ when using shortest paths to define height,
it was proved in \cite[Theorem~1.2]{stufler2022branching} that
the appropriately rescaled height process of a uniform galled tree with $n$
labeled leaves converges in distribution to the standard Brownian excursion:
this is, in fact, the key to the Gromov--Hausdorff convergence to the CRT.
Moreover, the proof is unchanged for $\square = \max$, as well as for
other notions of height.
Now, since the integral of the standard Brownian excursion
has the Airy distribution~\cite{flajolet2001analytic}, we immediately get
the convergence in distribution from~\eqref{eqCVSackinIndexGalledTrees};
the convergence of all moments then follows from the tail bounds
given in \cite[Theorem~1.3]{stufler2022branching}.
Note that this proof does not yield the numerical values of
the constants $\mu^{\min}$ and $\mu^{\max}$, but that it provides a probabilistic
interpretation for these constants (and that, in the case of galled trees, they
are readily recovered using the explicit description of the blowup construction
of uniform leaf-labeled galled trees given in~\cite{bienvenu2024b2}).

This approach has the advantage of working (1) not just for galled trees, but for
any blowup of a random tree that converges to the CRT for the Gromov--Hausdorff
topology (such as uniform leaf-labeled level-$k$ networks); (2) for
many notions of height, even in situations where the moments of the
corresponding extension of the Sackin index cannot be computed
explicitely. Moreover, thanks to the interpretation of the Sackin index as the
integral of the height process, it gives some intuition for
the fact that the limit distribution is the Airy distribution.
Finally, recent results on additive functionals of conditoned Galton--Watson
\cite{abraham2022global, fill2022sum}
trees suggest that a similar approach could be applicable not just to
extensions of the Sackin index, but also to extensions of other balance
indices.

\section{Concluding comments} \label{secConclusion}

Although non-exhaustive and biased by the author's taste when it comes to
the selection of techniques presented, this review should hopefully
provide a good overview of the state of the art concerning
the type of models of random phylogenetic networks that have been studied and
the type of results that have been obtained. We close this manuscript by
briefly mentioning two important lines of research.

First, since some of the tools presented here have only recently been
introduced to the field of phylogenetic networks, it is likely that these
have yet to be used to their full potential; in fact, one of the main
motivations for this review was to provide a non-technical introduction
to these tools -- in particular branching process techniques -- in the
hope that they become more widely used in the community. A first obvious step
in this direction, which would help identify the strengths and weaknesses
of each technique compared to better-established tools such as analytic
combinatorics, would be to see whether existing results can be reobtained with
simpler proofs using these tools. 

Second, as mentioned in the introduction, mathematically tractable
process-based models have an important role to play in applications; yet
as of today existing models are unsatisfactory from a biological point of
view. In particular, one of the main challenges seems to be developing tractable
models where phylogenetically distant lineages are less likely to
hybridize\,/\,experience horizontal gene transfer than closely related ones.
A first attempt at incorporating this in a tractable model was made in
\cite{bienvenu2024branching}, but this model has a tree-like structure almost
by definition, instead of as an emerging property -- which is not entirely
satisfying from a conceptual point of view. Thus, the problem
remains largely unaddressed; a discussion of this
topic can be found in \cite[Section~3]{couvert2023opening}.


\section*{Acknowledgements}

I thank three anonymous reviewers and the handling editor for helpful comments.
In particular, I thank a reviewer for pointing out that the convergence of the
height process of uniform level-$k$ networks formed by using only the leaves is
proved in~\cite{stufler2022branching}: this is the key to the short proof of
the convergence of the Sackin indices given in Section~\ref{secGHLimits}.  I
also thank Jean-Jil Duchamps for providing the code used to produce
Figure~\ref{figCRT}.

\addcontentsline{toc}{section}{References}
\bibliographystyle{abbrvnat}
\bibliography{biblio}

\begin{thebibliography}{64}
\providecommand{\natexlab}[1]{#1}
\providecommand{\url}[1]{\texttt{#1}}
\expandafter\ifx\csname urlstyle\endcsname\relax
  \providecommand{\doi}[1]{doi: #1}\else
  \providecommand{\doi}{doi: \begingroup \urlstyle{rm}\Url}\fi

\bibitem[Abraham and Delmas(2015)]{abraham2015introduction}
R.~Abraham and J.-F. Delmas.
\newblock An introduction to {Galton--Watson} trees and their local limits.
\newblock \emph{arXiv preprint}, 2015.
\newblock \arxiv{1506.05571}.

\bibitem[Abraham et~al.(2022)Abraham, Delmas, and Nassif]{abraham2022global}
R.~Abraham, J.-F. Delmas, and M.~Nassif.
\newblock Global regime for general additive functionals of conditioned
  {B}ienaym{\'e}--{G}alton--{W}atson trees.
\newblock \emph{Probability Theory and Related Fields}, 182:\penalty0 277--351,
  2022.
\newblock \doi{10.1007/s00440-021-01095-9}.

\bibitem[Aldous(1991{\natexlab{a}})]{aldous1991continuumI}
D.~Aldous.
\newblock The continuum random tree {I}.
\newblock \emph{The Annals of Probability}, 19\penalty0 (1):\penalty0 1--28,
  1991{\natexlab{a}}.
\newblock \doi{10.1214/aop/1176990534}.

\bibitem[Aldous(1991{\natexlab{b}})]{aldous1991continuumII}
D.~Aldous.
\newblock The continuum random tree {II}. {A}n overview.
\newblock pages 23--70. Cambridge University Press, 1991{\natexlab{b}}.
\newblock \doi{10.1017/CBO9780511662980.003}.

\bibitem[Aldous(1993)]{aldous1993continuum}
D.~Aldous.
\newblock The continuum random tree {III}.
\newblock \emph{The Annals of Probability}, 21\penalty0 (1):\penalty0 248--289,
  1993.
\newblock \doi{10.1214/aop/1176989404}.

\bibitem[Aldous and Steele(2004)]{aldous2004objective}
D.~Aldous and J.~M. Steele.
\newblock The objective method: Probabilistic combinatorial optimization.
\newblock In \emph{Probability on Discrete Structures. Encyclopaedia of
  Mathematical Sciences, volume~110}, Berlin, Heidelberg, 2004. Springer.
\newblock \doi{10.1007/978-3-662-09444-0_1}.

\bibitem[Barbour et~al.(1992)Barbour, Holst, and Janson]{barbour1992poisson}
A.~D. Barbour, L.~Holst, and S.~Janson.
\newblock \emph{Poisson Approximation}.
\newblock Oxford University Press, 1992.
\newblock ISBN 978-0-19-852235-5.

\bibitem[Baroni et~al.(2006)Baroni, Semple, and Steel]{baroni2006hybrids}
M.~Baroni, C.~Semple, and M.~Steel.
\newblock Hybrids in real time.
\newblock \emph{Systematic biology}, 55\penalty0 (1):\penalty0 46--56, 2006.
\newblock \doi{10.1080/10635150500431197}.

\bibitem[Benjamini and Schramm(2001)]{benjamini2001recurrence}
I.~Benjamini and O.~Schramm.
\newblock Recurrence of distributional limits of finite planar graphs.
\newblock \emph{Electronic Journal of Probability}, pages 1--13, 2001.
\newblock \doi{10.1214/EJP.v6-96}.

\bibitem[Bienvenu and Duchamps(2024)]{bienvenu2024branching}
F.~Bienvenu and J.-J. Duchamps.
\newblock A branching process with coalescence to model random phylogenetic
  networks.
\newblock \emph{Electronic Journal of Probability}, 29:\penalty0 1--48, 2024.
\newblock \doi{10.1214/24-EJP1088}.

\bibitem[Bienvenu and Steel(2024)]{bienvenu2024zero}
F.~Bienvenu and M.~Steel.
\newblock 0--1 laws for pattern occurrences in phylogenetic trees and networks.
\newblock \emph{Bulletin of Mathematical Biology}, 86\penalty0 (94), 2024.
\newblock \doi{10.1007/s11538-024-01316-x}.

\bibitem[Bienvenu et~al.(2021)Bienvenu, Cardona, and
  Scornavacca]{bienvenu2021revisiting}
F.~Bienvenu, G.~Cardona, and C.~Scornavacca.
\newblock Revisiting {S}hao and {S}okal's {$B_2$} index of phylogenetic
  balance.
\newblock \emph{Journal of Mathematical Biology}, 83\penalty0 (52):\penalty0
  1--43, 2021.
\newblock \doi{10.1007/s00285-021-01662-7}.

\bibitem[Bienvenu et~al.(2022)Bienvenu, Lambert, and
  Steel]{bienvenu2022combinatorial}
F.~Bienvenu, A.~Lambert, and M.~Steel.
\newblock Combinatorial and stochastic properties of ranked tree-child
  networks.
\newblock \emph{Random Structures \& Algorithms}, 60\penalty0 (4):\penalty0
  653--689, 2022.
\newblock \doi{doi.org/10.1002/rsa.21048}.

\bibitem[Bienvenu et~al.(2024)Bienvenu, Duchamps, Fuchs, and
  Yu]{bienvenu2024b2}
F.~Bienvenu, J.-J. Duchamps, M.~Fuchs, and T.-C. Yu.
\newblock The {$B_2$} index of galled trees.
\newblock \emph{arXiv preprint}, 2024.
\newblock \arxiv{2407.19454}.

\bibitem[Billingsley(1995)]{billingsley1995probability}
P.~Billingsley.
\newblock \emph{Probability and Measure}.
\newblock John Wiley \& Sons, 3rd edition, 1995.
\newblock ISBN 978-0-471-00710-4.

\bibitem[Bouvel et~al.(2020)Bouvel, Gambette, and Mansouri]{bouvel2020counting}
M.~Bouvel, P.~Gambette, and M.~Mansouri.
\newblock Counting phylogenetic networks of level 1 and 2.
\newblock \emph{Journal of mathematical biology}, 81\penalty0 (6):\penalty0
  1357--1395, 2020.
\newblock \doi{10.1007/s00285-020-01543-5}.

\bibitem[Caraceni et~al.(2022)Caraceni, Fuchs, and Yu]{caraceni2022bijections}
A.~Caraceni, M.~Fuchs, and G.-R. Yu.
\newblock Bijections for ranked tree-child networks.
\newblock \emph{Discrete Mathematics}, 345\penalty0 (9):\penalty0 112944, 2022.
\newblock \doi{10.1016/j.disc.2022.112944}.

\bibitem[Chang et~al.(2024{\natexlab{a}})Chang, Fuchs, Liu, Wallner, and
  Yu]{chang2024enumerative}
Y.-S. Chang, M.~Fuchs, H.~Liu, M.~Wallner, and G.-R. Yu.
\newblock Enumerative and distributional results for d-combining tree-child
  networks.
\newblock \emph{Advances in Applied Mathematics}, 157:\penalty0 102704,
  2024{\natexlab{a}}.
\newblock \doi{10.1016/j.aam.2024.102704}.

\bibitem[Chang et~al.(2024{\natexlab{b}})Chang, Fuchs, and Yu]{chang2024galled}
Y.-S. Chang, M.~Fuchs, and G.-R. Yu.
\newblock Galled tree-child networks.
\newblock \emph{arXiv preprint}, 2024{\natexlab{b}}.
\newblock \arxiv{2403.02923}.

\bibitem[Couvert et~al.(2023)Couvert, Bienvenu, Duchamps, Erard, Mir{\'o}~Pina,
  Schertzer, and Lambert]{couvert2023opening}
E.~Couvert, F.~Bienvenu, J.-J. Duchamps, A.~Erard, V.~Mir{\'o}~Pina,
  E.~Schertzer, and A.~Lambert.
\newblock Opening the species box: What microscopic models of speciation have
  to say about macroevolution.
\newblock \emph{bioRxiv preprint}, 2023.
\newblock \doi{10.1101/2023.11.09.564915}.

\bibitem[Duquesne(2003)]{duquesne2003limit}
T.~Duquesne.
\newblock A limit theorem for the contour process of conditioned
  {Galton--Watson} trees.
\newblock \emph{The Annals of Probability}, 31\penalty0 (2):\penalty0
  996--1027, 2003.
\newblock \doi{10.1214/aop/1048516543}.

\bibitem[Evans(2008)]{evans2008probability}
S.~N. Evans.
\newblock \emph{Probability and real trees}.
\newblock Springer, Berlin, Heidelberg, 2008.
\newblock \doi{10.1007/978-3-540-74798-7}.

\bibitem[Felsenstein(2003)]{felsenstein2004inferring}
J.~Felsenstein.
\newblock \emph{Inferring phylogenies}.
\newblock Sinauer Associates, 2003.
\newblock ISBN 978-0-87893-177-4.

\bibitem[Fill and Janson(2022)]{fill2022sum}
J.~A. Fill and S.~Janson.
\newblock The sum of powers of subtree sizes for conditioned {Galton--Watson}
  trees.
\newblock \emph{Electronic Journal of Probability}, 27:\penalty0 1--77, 2022.
\newblock \doi{10.1214/22-EJP831}.

\bibitem[Fischer et~al.(2023)Fischer, Herbst, Kersting, K{\"u}hn, and
  Wicke]{fischer2023tree}
M.~Fischer, L.~Herbst, S.~Kersting, A.~L. K{\"u}hn, and K.~Wicke.
\newblock \emph{Tree Balance Indices: A Comprehensive Survey}.
\newblock Springer, Cham, 2023.
\newblock \doi{10.1007/978-3-031-39800-1}.

\bibitem[Flajolet and Louchard(2001)]{flajolet2001analytic}
P.~Flajolet and G.~Louchard.
\newblock Analytic variations on the {A}iry distribution.
\newblock \emph{Algorithmica}, 31:\penalty0 361--377, 2001.
\newblock \doi{10.1007/s00453-001-0056-0}.

\bibitem[Flajolet and Sedgewick(2009)]{flajolet2009analytic}
P.~Flajolet and R.~Sedgewick.
\newblock \emph{Analytic Combinatorics}.
\newblock Cambridge University Press, 2009.
\newblock \doi{10.1017/CBO9780511801655}.

\bibitem[Fuchs and Gittenberger(2024)]{fuchs2024sackin}
M.~Fuchs and B.~Gittenberger.
\newblock Sackin indices for labeled and unlabeled classes of galled trees.
\newblock \emph{arXiv preprint}, 2024.
\newblock \arxiv{2407.13892}.

\bibitem[Fuchs et~al.(2022)Fuchs, Yu, and Zhang]{fuchs2022asymptotic}
M.~Fuchs, G.-R. Yu, and L.~Zhang.
\newblock Asymptotic enumeration and distributional properties of galled
  networks.
\newblock \emph{Journal of Combinatorial Theory, Series {A}}, 189:\penalty0
  105599, 2022.
\newblock \doi{10.1016/j.jcta.2022.105599}.

\bibitem[Fuchs et~al.(2024)Fuchs, Liu, and Yu]{fuchs2024limit}
M.~Fuchs, H.~Liu, and T.-C. Yu.
\newblock Limit theorems for patterns in ranked tree-child networks.
\newblock \emph{Random Structures \& Algorithms}, 64\penalty0 (1):\penalty0
  15--37, 2024.
\newblock \doi{10.1002/rsa.21177}.

\bibitem[Gusfield(2014)]{gusfield2014recombinatorics}
D.~Gusfield.
\newblock \emph{Re{C}ombinatorics: The Algorithmics and Combinatorics of
  Phylogenetic Networks with Recombination}.
\newblock The MIT Press, 2014.
\newblock \doi{10.7551/mitpress/9432.001.0001}.

\bibitem[Gusfield et~al.(2004{\natexlab{a}})Gusfield, Eddhu, and
  Langley]{gusfield2004fine}
D.~Gusfield, S.~Eddhu, and C.~Langley.
\newblock The fine structure of galls in phylogenetic networks.
\newblock \emph{{INFORMS} Journal on Computing}, 16\penalty0 (4):\penalty0
  459--469, 2004{\natexlab{a}}.
\newblock \doi{10.1287/ijoc.1040.0099}.

\bibitem[Gusfield et~al.(2004{\natexlab{b}})Gusfield, Eddhu, and
  Langley]{gusfield2004optimal}
D.~Gusfield, S.~Eddhu, and C.~Langley.
\newblock Optimal, efficient reconstruction of phylogenetic networks with
  constrained recombination.
\newblock \emph{Journal of bioinformatics and computational biology},
  2\penalty0 (01):\penalty0 173--213, 2004{\natexlab{b}}.
\newblock \doi{10.1142/S0219720004000521}.

\bibitem[Huson et~al.(2010)Huson, Rupp, and Scornavacca]{huson2010phylogenetic}
D.~H. Huson, R.~Rupp, and C.~Scornavacca.
\newblock \emph{Phylogenetic Networks: Concepts, Algorithms and Applications}.
\newblock Cambridge University Press, 2010.
\newblock \doi{10.1017/CBO9780511974076}.

\bibitem[Janson(2012)]{janson2012simply}
S.~Janson.
\newblock Simply generated trees, conditioned {G}alton--{W}atson trees, random
  allocations and condensation.
\newblock \emph{Probability Surveys}, 9:\penalty0 103--252, 2012.
\newblock \doi{10.1214/11-PS188}.

\bibitem[Janssen and Liu(2021)]{janssen2021comparing}
R.~Janssen and P.~Liu.
\newblock Comparing the topology of phylogenetic network generators.
\newblock \emph{Journal of bioinformatics and computational biology},
  19\penalty0 (06):\penalty0 2140012, 2021.
\newblock \doi{10.1142/S0219720021400126}.

\bibitem[Justison and Heath(2023)]{justison2023exploring}
J.~A. Justison and T.~A. Heath.
\newblock Exploring the distribution of phylogenetic networks generated under a
  birth-death-hybridization process.
\newblock \emph{Bulletin of the Society of Systematic Biologists}, 2\penalty0
  (3):\penalty0 1--22, 2023.
\newblock \doi{10.18061/bssb.v2i3.9285}.

\bibitem[Kersting et~al.(2024)Kersting, Wicke, and Fischer]{kersting2024tree}
S.~J. Kersting, K.~Wicke, and M.~Fischer.
\newblock Tree balance in phylogenetic models.
\newblock \emph{arXiv preprint}, 2024.
\newblock \arxiv{2406.05185}.

\bibitem[Khurana et~al.(2024)Khurana, Scheidwasser-Clow, Penn, Bhatt, and
  Duch{\^e}ne]{khurana2024limits}
M.~P. Khurana, N.~Scheidwasser-Clow, M.~J. Penn, S.~Bhatt, and D.~A.
  Duch{\^e}ne.
\newblock The limits of the constant-rate birth--death prior for phylogenetic
  tree topology inference.
\newblock \emph{Systematic Biology}, 73\penalty0 (1):\penalty0 235--246, 2024.
\newblock \doi{10.1093/sysbio/syad075}.

\bibitem[Knuth(1997)]{knuth1997art}
D.~E. Knuth.
\newblock \emph{The Art of Computer Programming: Fundamental Algorithms, Volume
  1}.
\newblock Addison-Wesley, 1997.
\newblock ISBN 978-0-201-89683-1.

\bibitem[Kong et~al.(2022)Kong, Pons, Kubatko, and Wicke]{kong2022classes}
S.~Kong, J.~C. Pons, L.~Kubatko, and K.~Wicke.
\newblock Classes of explicit phylogenetic networks and their biological and
  mathematical significance.
\newblock \emph{Journal of Mathematical Biology}, 84:\penalty0 47, 2022.
\newblock \doi{10.1007/s00285-022-01746-y}.

\bibitem[Kortchemski(2012)]{kortchemski2012invariance}
I.~Kortchemski.
\newblock Invariance principles for {{Galton}}--{{Watson}} trees conditioned on
  the number of leaves.
\newblock \emph{Stochastic Processes and their Applications}, 122\penalty0
  (9):\penalty0 3126--3172, 2012.
\newblock \doi{10.1016/j.spa.2012.05.013}.

\bibitem[Kurauskas(2022)]{kurauskas2022local}
V.~Kurauskas.
\newblock On local weak limit and subgraph counts for sparse random graphs.
\newblock \emph{Journal of Applied Probability}, 59\penalty0 (3):\penalty0
  755--776, 2022.
\newblock \doi{10.1017/jpr.2021.84}.

\bibitem[Lambert(2017)]{lambert2017probabilistic}
A.~Lambert.
\newblock Probabilistic models for the (sub)tree(s) of life.
\newblock \emph{Brazilian Journal of Probability and Statistics}, pages
  415--475, 2017.
\newblock \doi{10.1214/16-BJPS320}.

\bibitem[Lemant et~al.(2022)Lemant, Le~Sueur, Manojlovi{\'c}, and
  Noble]{lemant2022robust}
J.~Lemant, C.~Le~Sueur, V.~Manojlovi{\'c}, and R.~Noble.
\newblock Robust, universal tree balance indices.
\newblock \emph{Systematic biology}, 71\penalty0 (5):\penalty0 1210--1224,
  2022.
\newblock \doi{10.1093/sysbio/syac027}.

\bibitem[Miermont(2009)]{miermont2009tessellations}
G.~Miermont.
\newblock Tessellations of random maps of arbitrary genus.
\newblock \emph{Annales scientifiques de l'{\'E}cole normale sup{\'e}rieure},
  42\penalty0 (5):\penalty0 725--781, 2009.
\newblock \doi{10.24033/asens.2108}.

\bibitem[Morin and Moret(2006)]{morin2006netgen}
M.~M. Morin and B.~M. Moret.
\newblock {N\textsc{et}G\textsc{en}}: generating phylogenetic networks with
  diploid hybrids.
\newblock \emph{Bioinformatics}, 22\penalty0 (15):\penalty0 1921--1923, 2006.
\newblock \doi{10.1093/bioinformatics/btl191}.

\bibitem[Nakhleh et~al.(2002)Nakhleh, Sun, Warnow, Linder, Moret, and
  Tholse]{nakhleh2002towards}
L.~Nakhleh, J.~Sun, T.~Warnow, C.~R. Linder, B.~M. Moret, and A.~Tholse.
\newblock Towards the development of computational tools for evaluating
  phylogenetic network reconstruction methods.
\newblock In \emph{Biocomputing 2003}, pages 315--326. World Scientific, 2002.
\newblock \doi{10.1142/9789812776303_0030}.

\bibitem[Panagiotou et~al.(2023)Panagiotou, Ramzews, and
  Stufler]{panagiotou2023exact}
K.~Panagiotou, L.~Ramzews, and B.~Stufler.
\newblock Exact-size sampling of enriched trees in linear time.
\newblock \emph{SIAM Journal on Computing}, 52\penalty0 (5):\penalty0
  1097--1131, 2023.
\newblock \doi{10.1137/21M1459733}.

\bibitem[Pons et~al.(2019)Pons, Scornavacca, and Cardona]{pons2019generation}
J.~C. Pons, C.~Scornavacca, and G.~Cardona.
\newblock Generation of level-$k$ {LGT} networks.
\newblock \emph{{IEEE/ACM} Transactions on Computational Biology and
  Bioinformatics}, 17\penalty0 (1):\penalty0 158--164, 2019.
\newblock \doi{10.1109/TCBB.2019.2895344}.

\bibitem[Roch(2024)]{roch2024modern}
S.~Roch.
\newblock \emph{Modern Discrete Probability: An Essential Toolkit}.
\newblock Cambridge University Press, 2024.
\newblock \doi{10.1017/9781009305129}.

\bibitem[Ross(2011)]{ross2011fundamentals}
N.~Ross.
\newblock Fundamentals of {S}tein's method.
\newblock \emph{Probability Surveys}, 8:\penalty0 210 -- 293, 2011.
\newblock \doi{10.1214/11-PS182}.

\bibitem[Sackin(1972)]{sackin1972good}
M.~J. Sackin.
\newblock ``{G}ood'' and ``bad'' phenograms.
\newblock \emph{Systematic Biology}, 21\penalty0 (2):\penalty0 225--226, 1972.
\newblock \doi{10.1093/sysbio/21.2.225}.

\bibitem[Semple and Steel(2003)]{semple2003phylogenetics}
C.~Semple and M.~Steel.
\newblock \emph{Phylogenetics}, volume~24.
\newblock Oxford University Press, 2003.
\newblock \doi{10.1093/oso/9780198509424.001.0001}.

\bibitem[Shao and Sokal(1990)]{shao1990tree}
K.-T. Shao and R.~R. Sokal.
\newblock Tree balance.
\newblock \emph{Systematic Zoology}, 39\penalty0 (3):\penalty0 266--276, 1990.
\newblock \doi{10.2307/2992186}.

\bibitem[Steel(2016)]{steel2016phylogeny}
M.~Steel.
\newblock \emph{Phylogeny: Discrete and Random Processes in Evolution}.
\newblock SIAM, 2016.
\newblock \doi{10.1137/1.9781611974485}.

\bibitem[Steele(1994)]{steele1994lecam}
J.~M. Steele.
\newblock Le {C}am's inequality and {P}oisson approximations.
\newblock \emph{The American Mathematical Monthly}, 101\penalty0 (1):\penalty0
  48--54, 1994.
\newblock \doi{10.1080/00029890.1994.11996904}.

\bibitem[Stein(1972)]{stein1972bound}
C.~Stein.
\newblock A bound for the error in the normal approximation to the distribution
  of a sum of dependent random variables.
\newblock In \emph{Proceedings of the sixth {B}erkeley symposium on
  mathematical statistics and probability, volume~2: Probability theory}, pages
  583--603, 1972.

\bibitem[Stufler(2020)]{stufler2020limits}
B.~Stufler.
\newblock Limits of random tree-like discrete structures.
\newblock \emph{Probability Surveys}, 17:\penalty0 318--477, 2020.
\newblock \doi{10.1214/19-PS338}.

\bibitem[Stufler(2022)]{stufler2022branching}
B.~Stufler.
\newblock A branching process approach to {level-$k$} phylogenetic networks.
\newblock \emph{Random Structures \& Algorithms}, 61\penalty0 (2):\penalty0
  397--421, 2022.
\newblock \doi{10.1002/rsa.21065}.

\bibitem[{v}an~{d}er {H}ofstad(2024)]{vanderhofstad2024random}
R.~{v}an~{d}er {H}ofstad.
\newblock \emph{Random Graphs and Complex Networks}, volume~2.
\newblock Cambridge University Press, 2024.
\newblock \doi{10.1017/9781316795552}.

\bibitem[Zhang et~al.(2018)Zhang, Ogilvie, Drummond, and
  Stadler]{zhang2018bayesian}
C.~Zhang, H.~A. Ogilvie, A.~J. Drummond, and T.~Stadler.
\newblock Bayesian inference of species networks from multilocus sequence data.
\newblock \emph{Molecular biology and evolution}, 35\penalty0 (2):\penalty0
  504--517, 2018.
\newblock \doi{10.1093/molbev/msx307}.

\bibitem[Zhang(2019)]{zhang2019clusters}
L.~Zhang.
\newblock Clusters, trees, and phylogenetic network classes.
\newblock In \emph{Bioinformatics and Phylogenetics}, chapter~12, pages
  277--315. Springer, Cham, 2019.
\newblock \doi{10.1007/978-3-030-10837-3_12}.

\bibitem[Zhang(2022)]{zhang2022sackin}
L.~Zhang.
\newblock The {S}ackin index of simplex networks.
\newblock In \emph{{RECOMB} International Workshop on Comparative Genomics},
  pages 52--67. Springer, 2022.
\newblock \doi{10.1007/978-3-031-06220-9_4}.

\end{thebibliography}

\newpage

\phantomsection
\addcontentsline{toc}{section}{Supplementary Material}
\begin{center}
{\Large \bf Supplementary Material} \label{secSuppMat} \\[5ex]
\end{center}

In this supplementary material, we detail some of the calculations
for the birth-hybridization model with linear birth rate that are not given
in the main text.

\subsection*{Typical distance between the root and the leaves}

Consider the birth-hybridization process with hybridization probability
$\theta$, stopped upon reaching $n$ lineages (see
Box~\ref{boxLinearHybridization} from the main text).
Let $\gamma_n^{_\downarrow}$ be the random path obtained by starting from the
root and following the edges of the network until we reach a leaf, respecting
the orientation of the edges and choosing each outgoing uniformly at random and
independently of everything else when we reach a tree vertex;  and
let $|\gamma_n^{_\downarrow}| = \mathrm{length}(\gamma_n^{_\downarrow})$ denote
the number of tree edges of $\gamma_n^{_\downarrow}$.

Build the network and the random path $\gamma_n^{_\downarrow}$ jointly, in
forward time, as in the proof of point~(i) of
\cite[Theorem~5.2]{bienvenu2022combinatorial}: when there are $i=2$ lineages,
we have $|\gamma_2^{_\downarrow}| = 1$. Then, for every $i = 2, \ldots, n-1$, as
we go from $i$ to $i+1$ lineages,
\[
  |\gamma_{k+1}^{_\downarrow}| - |\gamma_{k}^{_\downarrow}| = X_i
\]
where $X_i$ is the event that the birth/hybridization creating the
$(i+1)$-th lineage involves the lineage through which
$\gamma_{k}^{_\downarrow}$ goes. Thus, we recover
Equation~\eqref{eqLengthGammaDown} from the main text:
\[
  |\gamma_{n}^{_\downarrow}| \;=\;
  1 \;+\; \sum_{i = 2}^{n-1} X_i \,,
\]
where $X_2, \ldots, X_{n-1}$ are independent.
Moreover, since once we fix a lineage:
\begin{itemize}
  \item one of the $i$ possible births involves this lineage;
  \item $i-1$ of the $\binom{i}{2}$ possible hybridizations involve it,
\end{itemize}
we have
\[
  \Prob{X_i = 1} \;=\; (1-\theta) \times \frac{1}{i} \;+\;
  \theta\times \frac{i-1}{\binom{i}{2}} \;=\;
  \frac{1 + \theta}{i} \,.
\]
Thus, setting $\ell^{_\downarrow}_n = 1 + (1+\theta) \sum_{i = 2}^{n-1} 1/i \sim
(1 + \theta) \log n$, for
$n$ large enough the
Stein--Chen bound~\eqref{eqSteinChenIndependent} gives
\[
  d_{\mathrm{TV}}\big(|\gamma_{n}^{_\downarrow}|, \, \mathrm{Poisson}(\ell^{_\downarrow}_n)\big) \;\leq\;
  \frac{1}{\ell^{_\downarrow}_n} \Big(1 + (1+\theta)^2 \sum_{i = 2}^{n-1} \frac{1}{i^2}
  \Big),
\]
which goes to zero as $n\to\infty$, proving the claims of
Box~\ref{boxLinearHybridization}.

Similarly, let $\gamma_n^{_\uparrow}$ be the random path 
obtained by choosing a leaf uniformly at random and then following the edges in
opposite direction until we reach the root, choosing each incoming edge
uniformly at random and independently of everything else when we reach a
reticulation. Again, we build the network and $\gamma_n^{_\uparrow}$ jointly
-- but this time in backwards time, using a coalescent process,
as in the proof of point~(ii) of \cite[Theorem~5.2]{bienvenu2022combinatorial}.
We get
\[
  |\gamma_n^{_\uparrow}| \;=\;
  1 \;+\; \sum_{i = 3}^n Y_i \,,
\]
where $Y_i$ is the indicator variable of the event that the lineage
through which $\gamma_n^{_\uparrow}$ goes is involved in the event taking us
from $i$ to $i-1$ lineages. Thus,
\[
  \Prob{Y_i = 1} \;=\;
  (1 - \theta) \times \frac{i-1}{\binom{i}{2}} \;+\; \theta\times
  \frac{3 \binom{i-1}{2}}{i \binom{i-1}{2}} \;=\; \frac{2+\theta}{i},
\]
since, once we have fixed the lineage through which $\gamma^{_\uparrow}_n$ goes:
\begin{itemize}
  \item $i-1$ of the $\binom{i}{2}$ coalescences that correspond to
    births  involve that lineage;
  \item $3\binom{i-1}{2}$
    of the $i\binom{i-1}{2}$ coalescences that correspond to
    hybridizations involve it.
\end{itemize}
The rest of the proof is the same as for $\gamma_{n}^{_\downarrow}$, but with $\ell^{_\uparrow}_n = 1 + (2 + \theta) \sum_{i = 3}^n 1/i$.

\subsection*{Number of reticulation leaves}

In this section, we pick up the study of the number of reticulation leaves from
where we left it at the end of Section~\ref{secSteinChen}, and only fill-in the
missing details.
We refer the reader to Section~2.2 of \cite{bienvenu2022combinatorial}
for more about the ``sticks-and-pearls'' construction of birth-hybridization
networks -- noting that the construction for uniform RTCNs immediately extends
to the birth-hybridization model with linear hybridization rate, using
the observation that conditional on their profile, networks produced by the
birth-hybridization model with linear hybridization rate are uniformly
distributed on the set of leaf-labeled RTCNs with that profile
(see Box~\ref{boxLinearHybridization} from the main text).

\enlargethispage{3ex}

Let us start by showing that the variables $X_3, \ldots, X_n$ are negatively
related, where we recall that
$X_i$ is the indicator variable of the event
\begin{align*}
  A_i \;&=\; \{\text{two pearls were placed at height $n-i+1$}\} \\
  &\qquad \cap\, \{\text{no unconditional pearl was placed on $S_i$}\} \,.
\end{align*}
For this, fix $i \in \Set{3, \ldots, n}$, set $I_i = \Set{3, \ldots, n} \setminus \{i\}$ and
consider the following coupling of $(X_j : j \in I_i)$ with a vector
$(\smash{X_j^{(i)}} : j \in I_i)$. First, place pearls on the sticks and
build the corresponding vector $(X_j : j \in I_i)$ as described in the main
text; then, to obtain 
$(\smash{X_j^{(i)}} : j \in I_i)$,
\begin{itemize}
  \item If only one pearl was placed at height $n-i+1$, say on $C^{(1)}_i$,
    then 
    choose a stick uniformly at random from $\Set{S_1, \ldots, S_{i-1}}
    \setminus \{\smash{C^{(1)}_i}\}$ and add a reticulation-inducing
    pearl on it. Note that this does not change $(X_j : j \in I_i)$.
  \item For each $k = i+1, \ldots, n$, if an unconditional pearl was placed on
    $S_i$ at height $n-k+1$, then move it to a stick chosen uniformly at random
    among the sticks $S_1, \ldots, S_{k-1}$ that have no reticulation-inducing
    pearl on them at height $n-k+1$. Note that if the pearl was moved to the
    stick $S_\ell$, this can only decrease~$X_\ell$.
\end{itemize}
After having made these modifications to the configuration of the pearls,
define $X^{(i)}_j$ to be the indicator variable of the event
$A_j$ for the modified configuration. With this construction,
\begin{mathlist}
\item  $(X_j^{(i)} : j \in I_i) \mathrel{\overset{d}{=}} ((X_j : j \in I_i) \mid
X_i = 1)$.
\item $X_j^{(i)} \leq X_j$ for each $j \in I_i$.
\end{mathlist}
Therefore,  the variables $X_3, \ldots, X_n$ are negatively
related.

Now, let us detail the computation of $\Var{W_n}$, where
we recall that $W_n = \sum_{i=3}^n X_i$. First, we have
\[
  \Var{W_n}   \;=\;
  \sum_{i=3}^n \Var{X_i} \;+\; 2 \sum_{i=3}^n \sum_{j = i+1}^n
  \Cov{X_i, X_j}\,,  
\]
where, by Equation~\eqref{eqProbXiEq1},
\[
  \Var{X_i} \;=\; \Prob{X_i = 1}\,\big(1 - \Prob{X_i = 1}\big)
  \;=\;\frac{\theta\,  (i-1)\, (n-1 -\theta(i-1))}{(n-1)^2}
\]
and, by Equations~\eqref{eqProbXiEq1} and~\eqref{eqProbXiXjEq1},
\begin{align*}
  \Cov{X_i, X_j} \;&=\;
  \Prob{X_i = 1, X_j = 1} - \Prob{X_i = 1} \Prob{X_j = 1} \\
  \;&=\; -\frac{\theta^2 (i-1)(n-j)}{(n-1)^2(n-2)} \,.
\end{align*}
Note that, as expected, this covariance is negative:
negative relation is a stronger notion of negative dependency than
pairwise negative correlation.

It then follows from a little algebra that
\[
  \Var{W_n}\;=\;
  \frac{\theta {\big((6 - 5 \theta) n^{3} + (10\, \theta - 12) n^{2}  +
  (7 \theta - 6) n + 12 - 24 \theta\big)}}{12 \, {\left(n - 1\right)}^{2}} \,, 
\]
which, as claimed in the main text, is asymptotically equivalent to
$\theta (6 - 5 \theta) n / 12$ as $n \to \infty$, and to 
$\theta n / 2$ when $\theta = o(1)$.

\end{document}